\documentclass[iop]{emulateapj}

\usepackage[fleqn]{amsmath}

\usepackage{color}

\newcommand{\rsun}{\ensuremath{R_\sun}}
\newcommand{\msun}{\ensuremath{M_\sun}}

\newcommand{\rstar}{\ensuremath{R_s}}

\newcommand{\rpl}{\ensuremath{R_{p}}}

\newcommand{\rjup}{\ensuremath{R_{\rm J}}}
\newcommand{\mjup}{\ensuremath{M_{\rm J}}}

\newcommand{\teff}{\ensuremath{T_{\rm eff}}}
\newcommand{\logg}{\ensuremath{\log g}}

\newcommand{\kms}{km~s$^{-1}$}

\newcommand{\ag}{\ensuremath{A_{\rm g}}}
\newcommand{\ab}{\ensuremath{A_{\rm B}}}
\newcommand{\eps}{\ensuremath{\varepsilon}}
\newcommand{\tn}{\ensuremath{T_{\rm N}}}
\newcommand{\td}{\ensuremath{T_{\rm D}}}
\newcommand{\tdeff}{\ensuremath{T_{\rm D, eff}}}

\newcommand{\tocc}{\ensuremath{T_{\rm occ}}}
\newcommand{\ttr}{\ensuremath{T_{\rm tr}}}

\newcommand{\dkep}{\ensuremath{D_{\ik}}}

\newcommand{\kep}{Kepler-13}
\newcommand{\kepb}{Kepler-13Ab}
\newcommand{\kepAb}{Kepler-13Ab}
\newcommand{\kepA}{Kepler-13A}
\newcommand{\kepB}{Kepler-13B}
\newcommand{\kepBB}{Kepler-13BB}
 
\newcommand{\ik}{{\it Kepler}}

\newcommand{\is}{{\it Spitzer}}

\newcommand{\ks}{$K_{s}$}

\newcommand{\mic}{\ensuremath {\rm \mu m}}

\newcommand{\sig}[1]{\ensuremath{#1\sigma}}

\newcommand{\figr}[1]{Fig.~\ref{fig:#1}}

\newcommand{\secr}[1]{Sec.~\ref{sec:#1}}

\newcommand{\eqr}[1]{Eq.~\ref{eq:#1}}
\newcommand{\eqsr}[2]{Eqs.~\ref{eq:#1}~and~\ref{eq:#2}}

\newcommand{\tabr}[1]{\mbox{Table~\ref{tab:#1}}}

\slugcomment{Accepted to ApJ on April 11 2014}

\shorttitle{\kepb\ atmospheric characterization}

\shortauthors{Shporer et al.}

\begin{document}

\title{Atmospheric Characterization of the Hot Jupiter \kepb}

\author{Avi Shporer\altaffilmark{1, 2, 17}, 
Joseph G.~O'Rourke\altaffilmark{1},
Heather A.~Knutson\altaffilmark{1},
Gyula M.~Szab\'o\altaffilmark{3, 4 ,5},
Ming Zhao\altaffilmark{6},
Adam Burrows\altaffilmark{7},
Jonathan Fortney\altaffilmark{8},
Eric Agol\altaffilmark{9},
Nicolas B.~Cowan\altaffilmark{10},
Jean-Michel Desert\altaffilmark{11},
Andrew W.~Howard\altaffilmark{12},
Howard Isaacson\altaffilmark{13},
Nikole A.~Lewis\altaffilmark{14, 15, 17},
Adam P.~Showman\altaffilmark{14},
Kamen O.~Todorov\altaffilmark{16}
} 
\altaffiltext{1}{Division of Geological and Planetary Sciences, California Institute of Technology, Pasadena, CA 91125, USA}
\altaffiltext{2}{Jet Propulsion Laboratory, California Institute of Technology, 4800 Oak Grove Drive, Pasadena, CA 91109, USA}
\altaffiltext{3}{ELTE Gothard Astrophysical Observatory, H-9704 Szombathely, Szent Imre herceg \'{u}t 112, Hungary}
\altaffiltext{4}{Konkoly Observatory, Research Centre of Astronomy and Earth Sciences, Hungarian Academy of Sciences, H-1121 Budapest, Konkoly Th. M. \'{u}t 15-17, Hungary}
\altaffiltext{5}{Gothard-Lend\"{u}let Research Team, H-9704 Szombathely, Szent Imre herceg \'{u}t 112, Hungary}
\altaffiltext{6}{Department of Astronomy and Astrophysics, 525 Davey Laboratory, The Pennsylvania State University, University Park, PA 16802, USA}
\altaffiltext{7}{Department of Astrophysical Sciences, Princeton University, Princeton, NJ 08544, USA}
\altaffiltext{8}{Department of Astronomy and Astrophysics, University of California, Santa Cruz, CA 95064, USA}
\altaffiltext{9}{Department of Astronomy, University of Washington, Seattle, WA 98195, USA}
\altaffiltext{10}{Center for Interdisciplinary Exploration and Research in Astrophysics and Department of Physics and Astronomy, Northwestern University, 2131 Tech Drive, Evanston, IL 60208, USA}
\altaffiltext{11}{CASA, Department of Astrophysical and Planetary Sciences, University of Colorado, 389-UCB, Boulder, CO 80309, USA}
\altaffiltext{12}{Institute for Astronomy, University of Hawaii, 2680 Woodlawn Drive, Honolulu, HI 96822, USA}
\altaffiltext{13}{University of California, Berkeley, CA 94720, USA}
\altaffiltext{14}{Department of Planetary Sciences and Lunar and Planetary Laboratory, The University of Arizona, Tucson, AZ 85721, USA}
\altaffiltext{15}{Department of Earth, Atmospheric and Planetary Sciences, Massachusetts Institute of Technology, Cambridge, MA 02139, USA}
\altaffiltext{16}{Institute for Astronomy, ETH Z\"{u}rich, Wolfgang-Pauli-Strasse 27, 8093 Z\"{u}rich, Switzerland}
\altaffiltext{17}{NASA Sagan Fellow}

%%%%%%%%%%%%%%%%%%%%%%%%%%%%%%%%%%%%%%
\begin{abstract}

\kepb\ (= KOI-13.01) is a unique transiting hot Jupiter. It is one of very few known short-period planets orbiting a hot A-type star, making it one of the hottest planets currently known. The availability of \ik\ data allows us to measure the planet's occultation (secondary eclipse) and phase curve in the optical, which we combine with occultations observed by warm \is\ at 4.5~\mic\ and 3.6~\mic\ and a ground-based occultation observation in the \ks\ band (2.1~\mic). We derive a day-side hemisphere temperature of 2,750$\pm$160 K as the effective temperature of a black body showing the same occultation depths. Comparing the occultation depths with one-dimensional planetary atmosphere models suggests the presence of an atmospheric temperature inversion. Our analysis shows evidence for a relatively high geometric albedo, \ag = $0.33^{+0.04}_{-0.06}$. While measured with a simplistic method, a high \ag\ is supported also by the fact that the one-dimensional atmosphere models underestimate the occultation depth in the optical. We use stellar spectra to determine the dilution, in the four wide bands where occultation was measured, due to the visual stellar binary companion 1\farcs15$\pm$0\farcs05 away. The revised stellar parameters measured using these spectra are combined with other measurements leading to revised planetary mass and radius estimates of $M_p$ = 4.94--8.09~\mjup\ and $R_p$ = 1.406$\pm$0.038~\rjup. Finally, we measure a \ik\ mid-occultation time that is 34.0$\pm$6.9 s earlier than expected based on the mid-transit time and the delay due to light travel time, and discuss possible scenarios.

\end{abstract}
%%%%%%%%%%%%%%%%%%%%%%%%%%%%%%%%%%%%%%

%% Keywords 

%%%%%%%%%%%%%%%%%%%%%%%%%%%%%%%%%%%%%%%%%
\section{Introduction}
\label{sec:intro}
%%%%%%%%%%%%%%%%%%%%%%%%%%%%%%%%%%%%%%%%%

The study of exoplanetary atmospheres is one of the most exciting aspects of the discovery of planets outside the Solar System. When the system is in a favorable edge-on geometric configuration the atmosphere of the unseen planet can be probed by measuring the decrease in observed flux during planetary transit (planet moves across the disk of its host star) or planetary occultation (secondary eclipse, when the planet moves behind the star), at different wavelengths. This approach favors large, hot, gas giant planets with large atmospheric scale heights, commonly known as hot Jupiters. This class of planets earns its name by having a radius about the radius of Jupiter while orbiting at short orbital periods, close-in to their host star. Tidal interaction is expected to lock (synchronize) the planet spin with the orbit, keeping the same planetary hemisphere constantly facing the star (a permanent day side) and the other hemisphere constantly facing away from the star (a permanent night side). Such planets do not exist in the Solar System, so only by probing the atmospheres of these distant worlds can we learn about atmospheric processes and atmospheric chemistry at such exotic environments.

The number of hot Jupiters whose atmospheres were studied using occultations is continuously rising and currently number in the several dozens. As the field transitions from the detailed study of individual objects to the characterization of a significant sample, several correlations, or patterns, are emerging. Several authors \citep[e.g.,][]{cowan11, perna12, perez13} have noticed that among the hot Jupiters, the hottest planets tend to have a low albedo and poor heat redistribution from the day side hemisphere to the night side hemisphere, pointing to a decreased advection efficiency. \cite{knutson10} noticed that planets with an inversion layer in their upper atmosphere, where temperature increases with decreasing pressure, tend to orbit chromospherically quiet (i.e.~non-active) stars, while planets with no inversion layer orbit chromospherically active stars. The inversion can be attributed to an absorber in the upper atmosphere that is being destroyed by UV radiation from chromospherically active stars \citep{knutson10}. Although, the occurrence of atmospheric inversions might also be related to atmospheric chemical composition, specifically the C to O elemental abundance ratio compared to the Solar composition value \citep{madhu11, madhu12}. Another interesting correlation involving chromospheric activity was identified by \cite{hartman10} who showed that planets with increased surface gravity tend to orbit stars with increased chromospheric activity. 

The patterns mentioned above are not fully explained. Gaining a better understanding of these patterns requires testing them with a larger sample, and studying planets at extreme environments and/or different characteristics, while observing over a wide range of wavelengths and obtaining a rough characterization of their spectrum. \kepb\ is such an extreme hot Jupiter, orbiting an A-type star every 1.76~days at a distance of only 0.034~AU. The close proximity to a hot, early-type star makes this planet one of the hottest currently known. With an irradiation at the planetary surface over 15,000 times that of Jupiter in the Solar System, the expected black body temperature of \kepb\ is up to over 3,000~K (for zero albedo and no heat redistribution from the day to night sides), comparable to the smallest stars, motivating the study of its atmosphere. Moreover, main-sequence A-type stars are inaccessible to spectroscopic radial velocity (RV) planet searches since their spectrum does not allow high precision RV measurements, making this a unique opportunity to study a planet in a short-period orbit around a main-sequence A-type star. The only other currently known hot Jupiter orbiting a bright A-type star is WASP-33b \citep{collier10, kovacs13}, although, that system is not in the \ik\ field and the host star's pulsating nature hampers the measurement of occultation depths. 

Here we carry out an atmospheric characterization of \kepb\ by measuring its occultation in four different wavelength bands, from the infrared (IR; \is/IRAC 4.5~\mic\ and 3.6~\mic), through the near-IR (NIR; \ks\ band), to the optical (\ik). We also analyze the \ik\ phase curve and obtain Keck/HIRES spectra that result in revised parameters for the objects in the system. We describe the analysis of our various data sets in \secr{dataanal}. In \secr{atm} we study the planet's atmosphere, and in \secr{dis} we discuss our results.

%===============================================================
\subsection{The \kep\ System}
\label{sec:kepler13}
%===============================================================

\begin{figure}
\begin{center}
\includegraphics[scale=0.40]{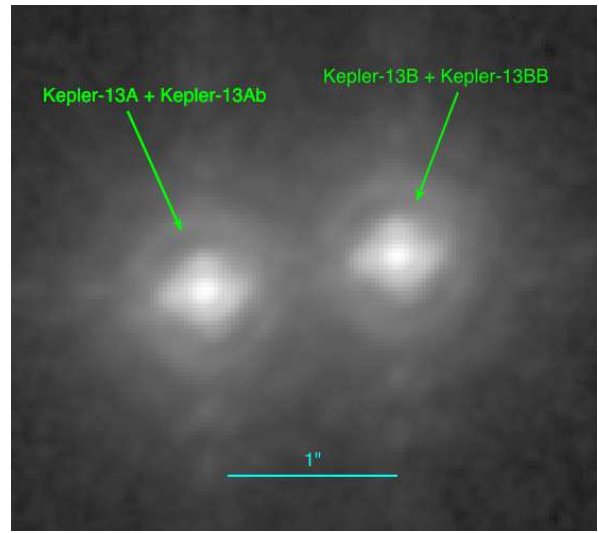}
\caption{\label{fig:fchart} 
High angular resolution adaptive optics imaging of the \kep\ system, obtained with P200/PHARO in the \ks\ band. North is up and East is to the left. The system is a visual binary, composed of two A-type stars at a sky-projected separation of 1\farcs15$\pm$0\farcs05. The brighter one, the primary, \kepA, is to the East (left), and is the planet host. The fainter component, the secondary, \kepB, is to the West (right) and is itself a stellar binary system, where the A-type star hosts a late-type star \citep{santerne12}. }
\end{center}
\end{figure}

The \kep\ system is a four body system, as far as we currently know. A high angular resolution image is shown in \figr{fchart}, taken from the publicly accessible \ik\ Community Follow-up Observing Program (CFOP) website\footnote{https://cfop.ipac.caltech.edu/}. The image was obtained in \ks\ (K short) band with the PHARO camera \citep{hayward01} and the adaptive optics system mounted on the Palomar 200 inch (5 m) Hale telescope (P200). The two bright components seen in \figr{fchart} are two A-type stars, where the brighter one, the primary (\kepA), hosts a transiting planet (\kepAb\footnote{In the literature it is occasionally referred to as simply Kepler-13b.}), while the fainter one, the secondary (\kepB), is orbited by a third star (\kepBB) of spectral type G or later \citep{santerne12}. The observed angular separation between the two A-type stars was measured to be 1\farcs12$\pm$0\farcs08 by \cite[][E.~Adams private communication]{adams12} and 1\farcs16$\pm$0\farcs06 by \cite{law13}, resulting in a weighted mean of 1\farcs15$\pm$0\farcs05. The distance to the system is 530~pc \citep{pickles10} with an uncertainty of 20\% (A.~Pickles, private communication), giving a sky-projected separation of $610 \pm 120$~AU. 

The availability of \ik\ data for a short-period planet transiting an A-type star in a bright ($V$~=~9.95~mag) hierarchical system makes it an interesting astrophysical laboratory. It is the first planet whose mass was estimated using photometric light curves \citep{shporer11, mazeh12, mislis12, esteves13, placek13}, and the first star-planet system where the star's obliquity was measured by modeling the asymmetric transit light curve due to stellar gravity darkening \citep{szabo11, barnes11}. In addition, orbital precession was also identified \citep{szabo12, szabo14}.

%%%%%%%%%%%%%%%%%%%%%%%%%%%%%%%%%%%%%%%%%
\section{Observations and Data Analysis}
\label{sec:dataanal}
%%%%%%%%%%%%%%%%%%%%%%%%%%%%%%%%%%%%%%%%%

All data used in this work is publicly available, either through dedicated archives of the relevant observatory, or by request from the authors. 

In all our photometric light curve analyses, from the IR (\is/IRAC), through the NIR (P200/WIRC/\ks), to the optical (\ik), we assume the measured flux is the combined flux from the two stars in the visual binary system (see \figr{fchart}), i.e., the two stars are fully blended together. We first analyze the data as is, without correcting for the dilution, and derive the {\it measured} occultation depth. Only then do we correct the measured depth, by estimating the amount of dilution, and derive the un-diluted, or {\it corrected} occultation depth. As shown in \secr{spec}, we estimate the dilution at each observed wavelength band by using Keck/HIRES resolved spectra of the two stars along with spectral models. 

Before moving to the description of the analysis of individual data sets we note that throughout this paper we have estimated the scatter in a given sample using the median absolute deviation from the median, also called median absolute deviation (MAD), defined as:
\begin{equation}
\rm MAD = median_j \left| X_j - median_i(X_i) \right| .
\end{equation}
This statistic is more robust than the root mean square as it is less sensitive to outliers, and hence provides a more reliable estimate of the standard deviation (StD) of the underlying distribution \citep[e.g.,][]{hoaglin83}. The StD, or \sig{1}, is calculated as:
\begin{equation}
\label{eq:mad2sigma}
\sigma = 1.4826 \cdot \rm  MAD .
\end{equation}

In addition, we note that when estimating the scatter in a given sample or when performing model fitting, we removed outliers iteratively while recalculating the scatter in each iteration until no outliers are left. The threshold, $n$, in units of $\sigma$, which is the  distance from the mean beyond which outliers were removed, was chosen to give an expectation value of less than 0.5 for the total number of outlier data points that exist in the sample, assuming a Gaussian distribution. This way, the threshold depends on the sample size, since as a sample grows larger so does the probability of data points having values further away from the median. This criteria can be written as the following inequality, which $n$ must satisfy given a sample of size $N$ with mean $\mu$ and standard deviation $\sigma$:
\begin{equation}
\label{eq:outlier}
1 - \frac{1}{2N} < \frac{1}{\sigma\sqrt{2\pi}} \int_{\mu-n\sigma}^{\mu+n\sigma} e^{-\frac{(x-\mu)^2}{2\sigma^2}} dx .
\end{equation}
For the sample sizes in this paper, ranging from about 1,000 up to about 50,000, the minimal value for $n$ ranges from 3.5 to 4.5. Removing data points in this way assures a negligible decrease in the measured scatter due to outlier removal, following a possible removal of data points belonging to the same Gaussian distribution but located at the distribution tails. We have also performed a visual inspection to verify that points identified as outliers do not appear to belong to the same Gaussian distribution.

%===============================================================
\subsection{\is\ $4.5~\mic$ Data}
\label{sec:spit2}
%===============================================================

Our analysis was done on \is\ Basic Calibrated Data produced by IRAC pipeline version S18.18.0 and obtained as part of Program ID 80219 (PI: H.~Knutson). Our \is/IRAC photometric and model fitting pipeline was implemented in Matlab, and we make our code publicly available\footnote{\url{http://gps.caltech.edu/$\sim$shporer/spitzerphot/}}.

We obtained a total of 3,973 full frame exposures, with an effective integration time of 4.4~s and a median cycle time of 7.2~s. The entire data set spans 7.83~h. 

Before analyzing the data we visually examined the 5\farcm2 $\times$ 5\farcm2 IRAC full frame images to verify that there are no bright stars other than the target within the intended photometric aperture or sky annulus. We also verified that the visual binary nature of the target does not affect its point spread function (PSF) shape, and that its width is comparable to that of other stars in the frame's field of view. The latter is expected since the IRAC pixel scale, of 1\farcs2 per pixel is comparable to the angular distance between the visual binary components, so the two stars are fully blended on \is/IRAC pixels.

%===============================================================
\subsubsection{Preprocessing}
\label{sec:prepro}
%===============================================================

First, we extracted the mid-exposure BJD of all images using the information in the image keyword headers. We used the BMJD\_OBS keyword timestamps, which is in UTC time, and added the leap seconds given by the difference between the ET\_OBS and UTCS\_OBS keyword timestamps. Next we added half the cycle time, given by the FRAMTIME header keyword, to get the mid exposure BJD:
\begin{eqnarray}
& \rm BJD = BMJD\_OBS\ +\ (ET\_OBS - UTCS\_OBS)\ +\ \nonumber \\
& \rm FRAMTIME/2\ +\  2,\!400,\!000.5 , 
\label{eq:bjd}
\end{eqnarray}
where ET\_OBS, UTCS\_OBS, and FRAMTIME were converted to units of days. \eqr{bjd} gives an approximate BJD\_TDB to better than one second \citep{eastman10}, which is sufficient for our needs. 

Using the target's median pixel position across all images, we extracted the 80 $\times$ 80 pixel region centered on the target from all 3,973 exposures. Those subframes were uploaded into a 3-dimensional (3D) matrix, whose dimensions are X pixel coordinate, Y pixel coordinate, and image number. 

Next, we removed the first 197 (= 5.0\%) images in the so-called \is\ ``ramp" --- which is a fast asymptotically-shaped instrumental increase in observed flux at the beginning of the observation. The cause for this fast initial ramp is not completely clear, but it may be due to telescope pointing settling or charge trapping. It has become common practice to trim out the initial part of the data, ranging from a few tens of minutes up to an hour or more, in order to minimize the scatter in the residuals from the best-fit model \citep[e.g.,][]{knutson12, todorov12, lewis13}. The data removed here includes the first 23~min (=~0.39~h) of the observation.

We have examined each pixel time series separately, along the image number dimension of the 3D matrix, and marked as bad pixels those deviating by more than \sig{5} from the series moving median. Such pixels were marked only in the specific frames where they showed a large deviation. No alignment between subframes was done prior to this step, as for this purpose the small pixel shifts between exposures, at the level of a few 0.01 pixel (see below), are negligible.

A final preprocessing step included conversion of pixel values from MJy~sr$^{-1}$ to electrons. This was done using the information in the FITS image header keywords, as described in the IRAC Instrument Handbook, Sec.~6.2\footnote{\url{http://irsa.ipac.caltech.edu/data/SPITZER/docs/irac/\\iracinstrumenthandbook/54/\#\_Toc296497452}}.

%===============================================================
\subsubsection{Aperture photometry}
%===============================================================

In each exposure the target's central pixel position was estimated by fitting a two dimensional Gaussian. We have also experimented with determining the central pixel position using the two dimensional center of mass (X and Y first moments), but achieved better results using a Gaussian fit, with a decreased scatter in the residuals from the best-fit model. The resulting scatter in central position X and Y coordinates is $\approx$0.01--0.02 pixels. \figr{xylc} shows the X and Y pixel position time series, where the familiar sawtooth behavior with a time scale of approximately 40~min is evident. This central position variation is attributed to an oscillation in \is\ pointing, which when combined with the non-uniform sensitivity across IRAC pixels results in similar oscillations in the measured flux.

We next calculated the target's flux in each image using aperture photometry with a circular aperture centered on the target, and summing over the sky-subtracted pixel values of pixels within the aperture. We used a time-varying aperture calculated using the ``noise pixel" parameter \cite[][see also IRAC Instrument Handbook Sec. 2.2.2\footnote{\url{http://irsa.ipac.caltech.edu/data/SPITZER/docs/irac/\\ iracinstrumenthandbook/5/}}]{mighell05, knutson12, lewis13}. In this method we first calculate:
\begin{equation}
\label{eq:noisepixel}
\tilde{\beta} = \frac{\left(\Sigma_i I_i\right)^2}{\Sigma_i I^2_i} ,
\end{equation}
where $I_i$ is the measured intensity in pixel $i$. This parameter gives the equivalent number of pixels that contribute to the point spread function, so is an approximation for the aperture radius squared. Hence, we take the aperture radius to be:
\begin{equation}
\label{eq:photrad}
r_j = c_0 + c_1\sqrt{\tilde{\beta_j}} ,
\end{equation}
where $r_j$ is the photometric aperture radius in image $j$, where the noise pixel parameter is $\tilde{\beta_j}$, and $c_0$ and $c_1$ are additive and multiplicative coefficients, respectively, that we optimize. Our best results were obtained for $c_1$ = 1.1, and $c_0$ = 0, meaning no additive factor. This approach resulted in a decreased scatter of the residuals from the best-fit model compared to using the same aperture for all frames.

The sky value was estimated as the median value of pixels within an annulus centered on the target, while iteratively removing \sig{5} outliers. The inner and outer annulus radii for which we obtained the best results were 10 and 30 pixels, respectively, although varying these values by a few pixels did not change the results significantly. Typically no more than five sky pixels were identified as outliers in each image.

Images where a bad pixel (see \secr{prepro}) was identified within the photometric aperture were ignored. This removed an additional 76 exposures from further analysis, which are 1.9\% of all exposures that together with the removal of the initial ramp amounts to 273 removed exposures, or 6.9\%, leaving 3,700 exposures. When estimating the median sky level bad pixels within the sky annulus were ignored. We also checked for exposures where the target's position was beyond \sig{4} away from the median position, and did not find any. 

Finally, we have median normalized the light curve measured in electrons, converting it into a relative flux light curve. We refer to this light curve as the {\it raw photometry} light curve, presented in \figr{xylc} bottom panel.

%===============================================================
\subsubsection{Post processing and model fitting}
%===============================================================

To model the raw photometry light curve we assumed the relative flux is a function of pixel position and time, $F(x,y,t)$, and that light curve variability is a consequence of three processes. The first is the planetary occultation, $O(t)$, the second is the non-uniform intra-pixel sensitivity, $M(x, y)$, and the third is a long-term time-dependent process, $T(t)$\footnote{This slowly varying temporal component is sometimes also referred to as a ramp in the literature, not to be confused with the fast initial ramp that is trimmed out (see \secr{prepro}).}. Therefore our model is:
\begin{equation}
\label{eq:lcmodel}
F(x,y,t) = O(t) \times M(x,y) \times T(t) .
\end{equation}
For the occultation model, $O(t)$, we used a model based on the \cite{mandel02} transit model, with the occultation depth as an additional parameter to account for the planet's luminosity. Also, the stellar limb darkening is irrelevant here, and we assume uniform surface brightness across the planet's disk. Our model assumes a circular orbit, following the expectation of complete circularization of such close-in planetary orbits \citep[e.g.,][]{mazeh08}. Only the occultation depth was allowed to vary freely. The ephemeris was fixed to the one derived here (see \secr{kepler_occ}) while the rest of the light curve parameters were fixed to the known values \citep{barnes11} as there is not enough signal-to-noise in the \is\ data to efficiently constrain them. For the CCD position-dependent and time-dependent components we used polynomials:
\begin{align}
\label{eq:pospoly}
& \hspace{-10mm}  M(x,y) = M_0 + M_{1,x}x + M_{2,x}x^2 + M_{1,y}y + M_{2,y}y^2 , \\
\label{eq:timepoly}
& \hspace{-10mm}  T(t) = 1 + T_1t ,
\end{align}
where adding higher order and/or mixed terms (in \eqr{pospoly}) did not decrease the uncertainty on the fitted occultation depth or improve the scatter of the residuals from the best-fit model.

The above model includes seven free parameters: the occultation depth and six polynomial coefficients, which we fitted using the Monte Carlo Markov Chain (MCMC) algorithm as described in \cite{shporer09}. We ran a total of 5 chains, consisting of $10^6$ steps each, resulting in Gaussian distributions for the fitted parameters, after ignoring the initial 20\% of each chain. We took the distributions median to be the best-fit value and the values at 84.13 and 15.87 percentiles to be the +\sig{1} and -\sig{1} confidence limits, respectively. We compared the errors from the MCMC approach with errors estimated using the ``residual permutation" method, referred to also as the ``prayer bead" method \citep[e.g.,][]{gillon07, carter09}, and adopted the larger of the two errors for each fit parameter. In the prayer bead method we cyclicly shifted the residuals with respect to time and refitted. This way each fit is affected by the same correlated noise that may exist in the original data. This approach results in a distribution for each fitted parameter where we take the values at 84.13 and 15.87 percentiles to be the +\sig{1} and -\sig{1} confidence limits. We verified that the distributions' median were indistinguishable from the initial fitted values. 

The resulting occultation depth is listed in \tabr{res} along with the scatter of the residuals, which is 2.1\% larger than the expected Poisson noise. \tabr{res} also lists the scatter while using one minute bins, also referred to as the photometric noise rate (PNR; e.g., \citealt{shporer09, fulton11}). The latter allows an easy comparison between light curves obtained with different instruments, with different exposure times and cycle times. 

We have experimented with other models and fitting methods, to confirm our results. First, we replaced the \cite{mandel02} model with a trapezoid model, since they differ only during the occultation ingress and egress phase and the difference is up to only a few 10$^{-5}$ in relative flux. The results we got using this model are the same as when using the original one. Another approach we tried, motivated by the sinusoidal shape of the position-dependent flux modulations (see \figr{xylc}), was adding sinusoidal harmonics to the long-term time-dependent process ($T(t)$)  while removing the $M(x,y)$ component from the model (see \eqr{lcmodel}). Combined with the trapezoid model this allows a linear least squares fitting of the entire light curve model using a single matrix inversion operation, making it CPU efficient. However, this approach gave poorer results than the original one (larger residuals scatter with clear systematic features). A possible reason for this is that the position-dependent flux modulations are not well approximated by sinusoids.

The 4.5~\mic\ occultation light curve is listed in \tabr{lc45m}, and shown in \figr{lc}, after removing the positional ($M(x,y)$) and long-term temporal ($T(t)$) components. In \figr{bin} we plot the scatter of the residuals for a range of bin sizes. Although the scatter decreases with the inverse square root of the bin width as expected for Poisson noise, there does seem to be a correlated noise component at 1--10 minutes whose source is unknown.

\begin{deluxetable*}{cccccc} 
\tablecaption{\label{tab:lc45m} {\it Spitzer} 4.5 $\rm \mu m$ light curve}  
\tablewidth{0pt} 
\tablehead{  
\colhead{Time} & \colhead{X} & \colhead{Y} & \colhead{Raw Rel.~Flux\tablenotemark{a}} & \colhead{Rel.~Flux Error} & \colhead{Detrended Rel.~Flux\tablenotemark{b}}   \\ 
\colhead{BJD} & \colhead{pixel} & \colhead{pixel} & \colhead{} & \colhead{} & \colhead{}   \\ 
} 
\startdata 
2455795.4900269  &  127.04045  &  128.57439  &  1.002220  &  0.003385  &  1.003386  \\
2455795.4901103  &  127.03541  &  128.56319  &  1.002003  &  0.003384  &  1.002439  \\
2455795.4901936  &  127.03748  &  128.56890  &  1.000259  &  0.003381  &  1.001002  \\
2455795.4902723  &  127.03763  &  128.56315  &  0.999927  &  0.003381  &  1.000507  \\
2455795.4903556  &  127.04253  &  128.56182  &  0.999790  &  0.003381  &  1.000691  \\
2455795.4904389  &  127.03489  &  128.55481  &  0.997916  &  0.003378  &  0.998151  \\
2455795.4905177  &  127.04364  &  128.55421  &  0.993622  &  0.003370  &  0.994462  \\
2455795.4906010  &  127.04522  &  128.56129  &  0.999825  &  0.003381  &  1.000926  \\
2455795.4906844  &  127.04823  &  128.56173  &  0.999433  &  0.003380  &  1.000797  \\
2455795.4907676  &  127.05526  &  128.58094  &  0.997450  &  0.003377  &  1.000194  \\
\enddata 
\tablenotetext{a}{Raw relative flux, without correcting for the intra-pixel sensitivity variations.} 
\tablenotetext{b}{Detrended relative flux, after removing the intra-pixel sensitivity variations.} 
\tablecomments{Columns include, from left to right: Mid-exposure BJD, X coordinate, Y coordinate, Raw relative flux, relative flux error, and detrended light curve after removing all variabilities except the occultation. The table in its entirety is available on-line.}
\end{deluxetable*}

%===============================================================
\subsection{\is\ $3.6~\mic$ Data}
\label{sec:spit1}
%===============================================================

Our approach in analyzing these data was largely similar to that of the 4.5~\mic\ analysis. Basic Calibrated Data files were produced by IRAC pipeline version S19.1.0, and data were obtained as part of Program ID 80219 (PI: H.~Knutson). We obtained 14,144 individual exposures spanning 7.94~h, with an exposure time of 1.92 s and a median cycle time of 2.00 s. 

We ignored the first 95~min (1.58~h) of data due to the fast initial ramp, corresponding to 2,809 exposures or 19.9\% of all exposures. This ramp is longer than the one we removed in the 4.5 \mic\ data, consistent with other studies that found a longer initial ramp in this band than in 4.5 \mic\ \citep[e.g.,][]{knutson12, lewis13}. We then rejected additional 251 exposures, or 1.8\%, with a bad pixel identified within the photometric aperture. Therefore, the number of exposures used in the analysis is 11,084, spanning 6.36~h.

\begin{figure*}[t]
\begin{center}
\includegraphics[width=17.5cm]{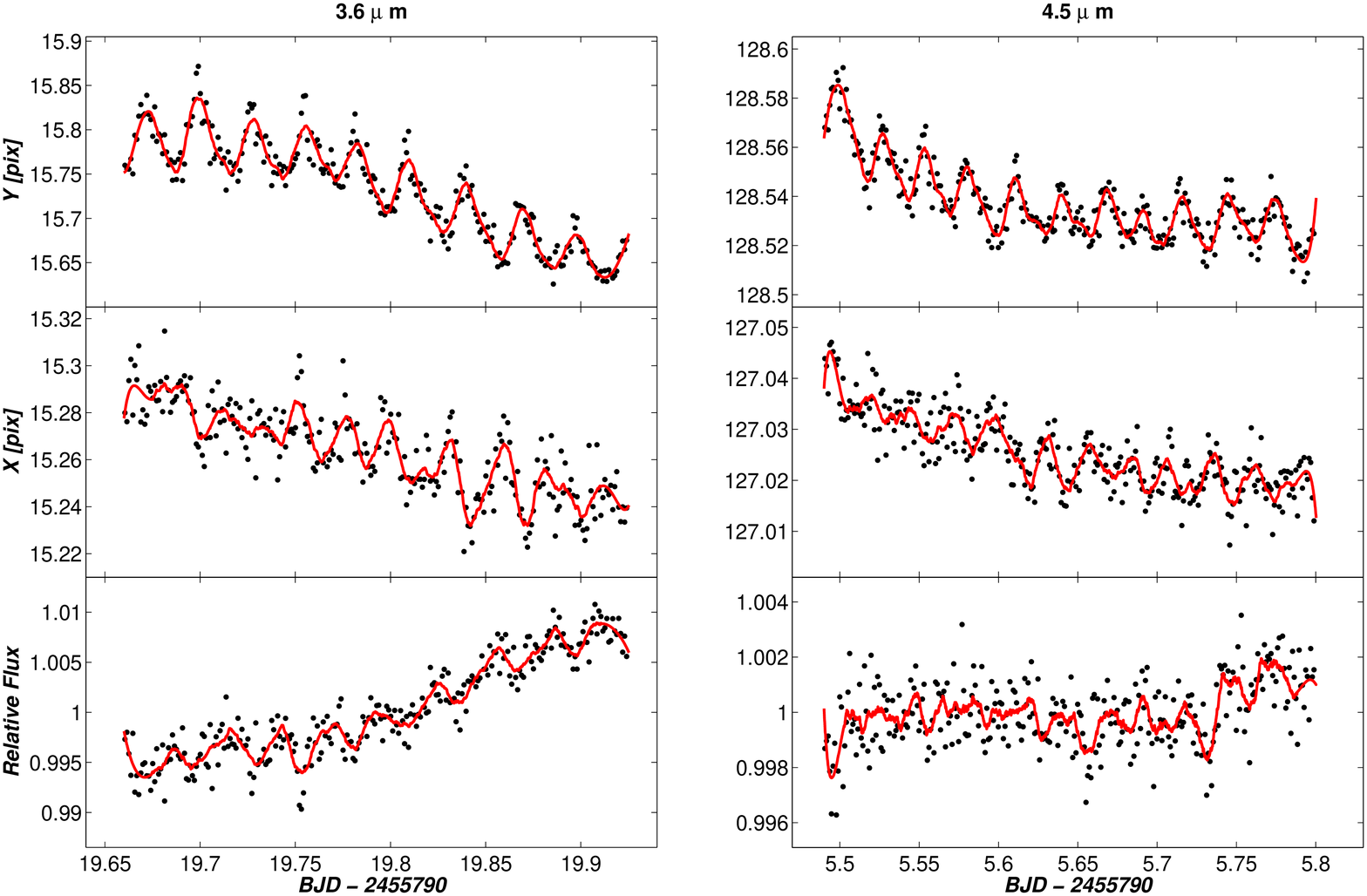}
\caption{\label{fig:xylc} 
\is\ pixel position and raw photometry binned time series, in 3.6~\mic (left) and 4.5~\mic (right). Top panels show the Y pixel position, middle panels the X pixel position, and bottom panels the raw photometry. The overplotted solid red line is a Savitzky-Golay smoothing, showing the familiar sawtooth pattern, with a time scale of 40~min.
}
\end{center}
\end{figure*}

We used a two dimensional Gaussian fit to determine the central pixel position, with a resulting scatter of 0.03~pixel and 0.07~pixel in the X and Y coordinates, respectively. The sawtooth pattern is seen in both coordinates time series, shown in \figr{xylc}. 

We determined the aperture used in each frame based on the noise pixel parameter, $\tilde\beta$ (see \eqr{noisepixel}), where the aperture radius was taken to be $1.0 \times \sqrt{\tilde\beta}$, meaning $c_1$ = 1.0, and $c_0$ = 0 (see \eqr{photrad}). The sky value was taken to be the median of pixels within an annulus between radii of 10 and 20 pixels while ignoring bad pixels.

Here our fitted model did not include a time component, meaning $T(t)$ was taken to be unity in \eqr{lcmodel}, since when including a linear component as a function of time ($T_1$ in \eqr{timepoly}) the fitted coefficient was consistent with zero and removing it did not affect any of the results including the residual's scatter and occultation depth. Therefore, the total number of fitted parameters was six: the occultation depth and five polynomial coefficients.

\begin{figure}
\begin{center}
\includegraphics[scale=0.50]{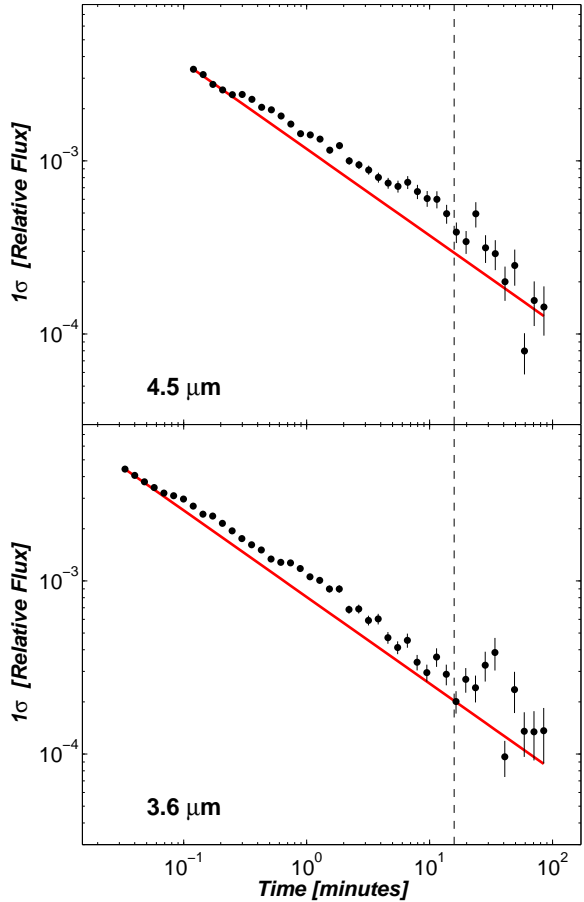}
\caption{\label{fig:bin} 
Light curve residuals scatter vs.~bin width (black filled circles with error bars), in log-log scale, for the \is\ 3.6~\mic\ light curve (bottom) and the 4.5~\mic\ light curve (top). The solid red line shows the expected decrease in scatter according to Poisson statistics, and the vertical dashed line marks ingress/egress duration.
}
\end{center}
\end{figure}

The resulting occultation depth is listed in \tabr{res}, along with the residuals scatter and PNR. The scatter is 18.7\% larger than the expected Poisson noise. Although this is more than for the 4.5 \mic\ data it is comparable to the excess noise level above the Poisson noise level found in other studies \citep[e.g.,][]{knutson12, orourke14}. The fitted occultation light curve is listed in \tabr{lc36m} and plotted in \figr{lc}, and the residuals scatter vs.~bin size is plotted in \figr{bin}.

\begin{deluxetable*}{cccccc} 
\tablecaption{\label{tab:lc36m} {\it Spitzer} 3.6 $\rm \mu m$ light curve}  
\tablewidth{0pt} 
\tablehead{  
\colhead{Time} & \colhead{X} & \colhead{Y} & \colhead{Raw Rel.~Flux\tablenotemark{a}} & \colhead{Rel.~Flux Error} & \colhead{Detrended Rel.~Flux\tablenotemark{b}}   \\ 
\colhead{BJD} & \colhead{pixel} & \colhead{pixel} & \colhead{} & \colhead{} & \colhead{}   \\ 
} 
\startdata 
2455809.6600049  &   15.28237  &   15.81692  &  0.986185  &  0.004432  &  0.991893  \\
2455809.6600281  &   15.26722  &   15.79148  &  0.994718  &  0.004451  &  0.997973  \\
2455809.6600512  &   15.27482  &   15.78554  &  0.995372  &  0.004453  &  0.999090  \\
2455809.6600744  &   15.28386  &   15.78273  &  0.996489  &  0.004455  &  1.000945  \\
2455809.6600975  &   15.28723  &   15.78937  &  1.000421  &  0.004464  &  1.005526  \\
2455809.6601207  &   15.27445  &   15.77484  &  0.993759  &  0.004449  &  0.996927  \\
2455809.6601438  &   15.27659  &   15.76911  &  0.998592  &  0.004460  &  1.001693  \\
2455809.6602072  &   15.26584  &   15.74978  &  0.992963  &  0.004447  &  0.993996  \\
2455809.6602304  &   15.25616  &   15.76161  &  0.998753  &  0.004460  &  0.999500  \\
2455809.6602535  &   15.26450  &   15.76773  &  0.993252  &  0.004448  &  0.995111  \\ 
\enddata
\tablenotetext{a}{Raw relative flux, without correcting for the intra-pixel sensitivity variations.}
\tablenotetext{b}{Detrended relative flux, after removing the intra-pixel sensitivity variations.}
\tablecomments{Columns include, from left to right: Mid-exposure BJD, X coordinate, Y coordinate, Raw relative flux, relative flux error, and detrended light curve after removing all variabilities except the occultation. The table in its entirety is available on-line.}
\end{deluxetable*} 

%===============================================================
\subsection{Palomar 200 inch WIRC \ks\ Data}
\label{sec:wircks}
%===============================================================

We observed an occultation of \kepb\ in the \ks\ band on UT 2012 August 28 with the wide field infrared camera \citep[WIRC;][]{wilson03}, mounted on the P200. The WIRC instrument has a 2048 $\times$ 2048 pixel detector, a pixel scale of 0\farcs2487 per pixel, and a field of view of 8\farcm7 $\times$ 8\farcm7. Observation of the target field began at 03:38:04.8 UTC and ended 6.57 h later while obtaining a total of 1,123 images. The first image was rejected because the telescope guiding had not yet stabilized, and so were the last three images due to increased background level as the night ended. The airmass ranged from 1.03 -- 2.15 during the observation. The seeing varied throughout the night but remained $\lesssim1\arcsec$. We began with an exposure time of 9 s, but decreased the exposure time incrementally to 8 s, 7 s, and finally 6 s throughout the night to keep pixel counts in the detector's linear regime, well below saturation. We also defocused the telescope to a PSF FWHM of $\sim$2$\farcs5$--$3\farcs$0 to avoid saturation and systematic errors resulting from variations in intra-pixel sensitivity. As a result \kepA\ and \kepB\ were completely blended in all images, and we treated the target as a single source in our photometric analysis. To minimize systematics related to imperfect flat-fielding and inter-pixel variations in the detector, we did not dither the telescope.

Data reduction and analysis was done using a pipeline developed specifically for WIRC data, described in more detail in \cite{orourke14}, and we give only a short description here. Images were dark-subtracted and flat-fielded using the median of 18 normalized twilight flats as a single master flat field. We selected nine reference stars with median fluxes ranging from $\sim$0.09 -- 1.80 times that of \kep. Two bright stars in the field were ignored as their fluxes consistently saturated the detector or exceeded the linearity regime. We performed aperture photometry on each star using circular apertures with fixed radii and determined the sky background level using an annulus centered on the star's position. Aperture radius and sky annulus inner and outer radii were optimized to minimize the scatter in our final light curve model fit, yielding an aperture of 20.0 pixels and sky annulus inner and outer radii of 25.0 and 55.0 pixels respectively. At this point we excluded another 16 images from the analysis because either pixel counts in the photometric apertures exceeded the detector's linearity regime or the total flux of the target or one of the reference stars varied by more than \sig{3} from the median value in the adjacent 20 frames in the time series.

For each measurement we calculated the mean of the nine reference stars and derived a single reference light curve (using the median or flux-weighted mean produced inferior results in the eventual fit). We divided the light curve for \kep\ by the reference light curve. Then, we fitted this normalized light curve simultaneously with a linear trend with time and a model for the occultation. As done for the \is\ data, we used an occultation model based on \citet{mandel02} while allowing only the occultation depth to vary freely and keeping all other occultation light curve parameters fixed. We used the ephemeris derived here (see \secr{kepler_occ}) and adopt the rest of the model parameter values from the literature \citep{barnes11}. We fitted the light curve model using the MCMC algorithm and also used the prayer bead method where the estimated uncertainties were found to be in good agreement.

\begin{figure}
\begin{center}
\includegraphics[scale=0.50]{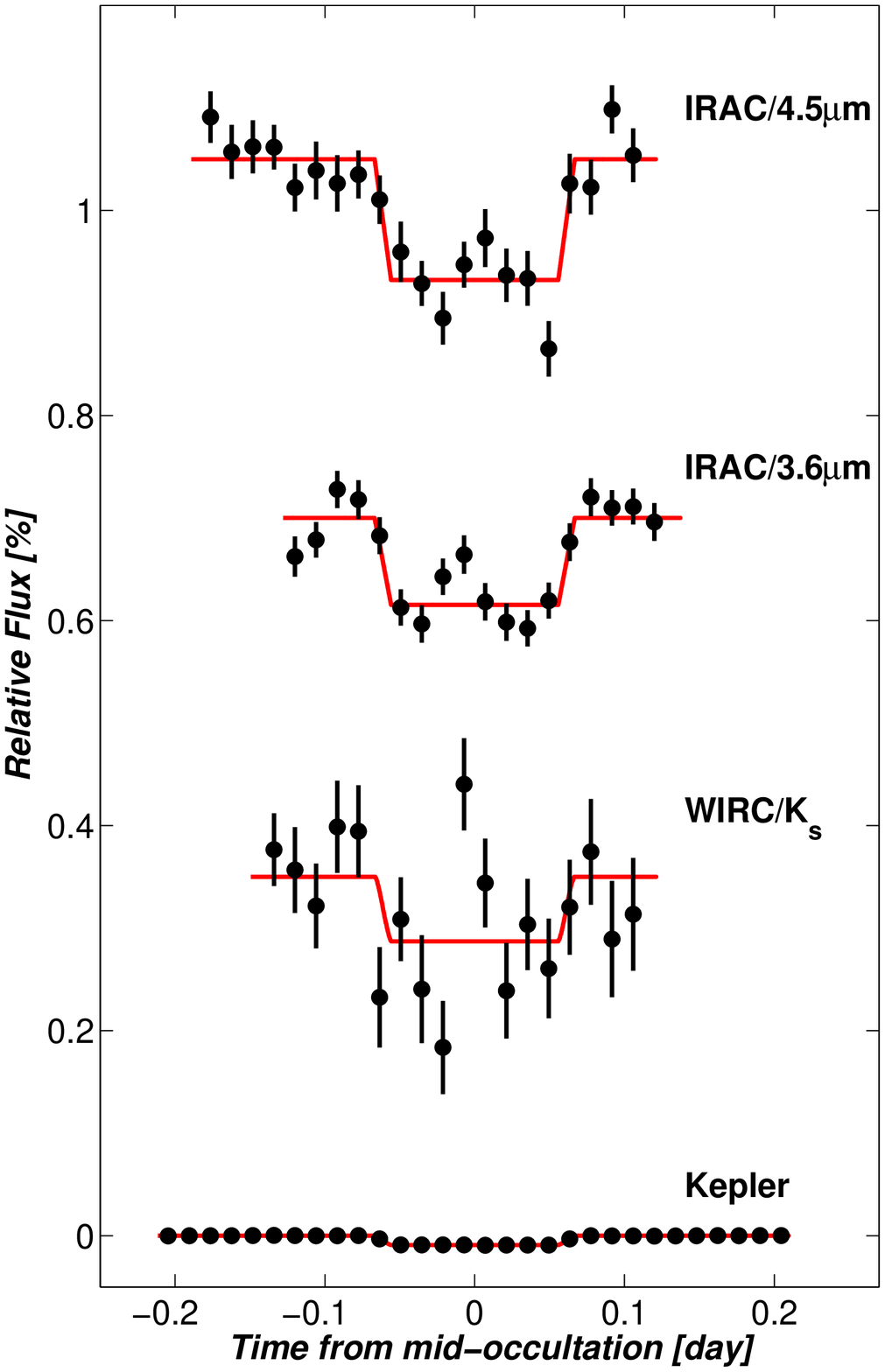}
\caption{\label{fig:lc} 
Binned occultation light curves in (from top to bottom) 4.5~\mic\ and 3.6~\mic\ (\is), \ks\ band (WIRC), and optical (\ik). The solid red line is the occultation model. Light curves are plotted in relative flux vs.~ time from mid-occultation and are shifted vertically for visibility purpose. The \ik\ occultation depth is too shallow to be seen on this scale, and the light curve is presented here just for reference while it is shown in detail in \figr{occkep}.
}
\end{center}
\end{figure}

We derived an occultation depth of 0.063~$\pm$~0.026\%. The residuals scatter for the best-fit solution is 0.389\%, a factor of 3.56 larger than the Poisson noise limit of 0.109\%. Such noise levels are similar to those obtained for other WIRC data sets \citep{zhao12a, zhao12b, orourke14}. The WIRC/\ks\ occultation light curve is shown in \figr{lc} and listed in \tabr{lcwircks}.

\begin{deluxetable*}{cccc} 
\tablecaption{\label{tab:lcwircks} WIRC K$_{\rm s}$ light curve}  
\tablewidth{0pt} 
\tablehead{  
\colhead{Time} & \colhead{Rel.~Flux} & \colhead{Rel.~Flux Error} & \colhead{Detrended Rel.~Flux}   \\ 
\colhead{BJD} & \colhead{} & \colhead{} & \colhead{}   \\ 
} 
\startdata 
 2456167.646833  &   1.00189  &   0.00389  &   0.99970  \\
 2456167.647051  &   1.00075  &   0.00389  &   0.99856  \\
 2456167.647264  &   1.00443  &   0.00389  &   1.00224  \\
 2456167.647480  &   1.00513  &   0.00389  &   1.00294  \\
 2456167.647907  &   0.99883  &   0.00389  &   0.99667  \\
 2456167.648214  &   1.00299  &   0.00389  &   1.00083  \\
 2456167.648473  &   1.00050  &   0.00389  &   0.99835  \\
 2456167.648735  &   0.99815  &   0.00389  &   0.99601  \\
 2456167.648994  &   0.99983  &   0.00389  &   0.99770  \\
 2456167.649254  &   1.00573  &   0.00389  &   1.00359  \\
\enddata 
\tablecomments{Columns include, from left to right: Mid-exposure BJD, relative flux (without correcting for the long term trend), relative flux error, and detrended light curve after removing the long term temporal trend. The table in its entirety is available on-line.}
\end{deluxetable*} 

%===============================================================
\subsection{\ik\ Occultation Data}
\label{sec:kepler_occ}
%===============================================================

\ik\ data allow a high precision measurement of \kepb\ occultation in the optical. This is thanks to the high precision photometry, the many occultation events observed, and the availability of a large amount of short cadence data, which we used for ten quarters: Q2--Q3 and  Q7--Q14. 

We processed the \ik\ short cadence occultation data by first going through each occultation event and normalizing it using a 2nd degree polynomial fitted to the out of occultation data. Using a 3rd degree polynomial did not change the results. For this normalization we used data segments centered on the occultation and spanning a total of three times the occultation itself, including out of occultation data before ingress and after egress spanning the same duration as the occultation. We considered only full events, meaning where the entire occultation event was observed including both pre-ingress and post-egress data. Polynomial fitting was done by iteratively removing \sig{4} outliers until none were identified (there were approximately 600 data points in each individual occultation light curve segment). We also removed occultation events occurring at times where the data showed strong trends, for example, near spacecraft safe modes\footnote{\url{http://archive.stsci.edu/kepler/manuals/Data\_Characteristics.pdf}}. We were left with 453 viable occultation events out of the 522 events that occurred during the above 10 quarters, which are 87\% of the events. For the majority of the other 13\% events no data were obtained at all as they happened during breaks in data collection due to data downloads, quarterly spacecraft rotations, and safe modes.

Next, we fitted the data using an occultation light curve model based on the \cite{mandel02} transit model as done for the \is\ and WIRC data. Despite the high quality of the \ik\ short cadence occultation data, it cannot resolve the degeneracy between some of the model parameters, specifically the planet to star radii ratio, $R_p/R_s$, the orbital semi-major axis normalized by the stellar radius, $a/R_s$, the occultation impact parameter, $b$, and occultation depth, $D_{\ik}$. Therefore, we used Gaussian priors on three of the parameters: $R_p/R_s$, $a/R_s$, and $b$, and fitted as free parameters the other three model parameters, including the orbital period, $P$, a specific mid-occultation time, \tocc, and the occultation depth. Gaussian priors were taken from \cite{barnes11} where fitting the transit light curve resolved the three parameters $R_p/R_s$, $a/R_s$, and $b$. 

After obtaining a preliminary fitted model in the manner described below, we repeated the fit while iteratively removing outliers beyond \sig{5.5}, rejecting 0.27\% of the 278,901 data points. Visual inspection showed that increasing the outlier threshold did not remove some clear outliers, while decreasing it removed data points that did not seem to be outliers.

We carried out model fitting using the MCMC algorithm. We ran a total of 5 chains, consisting of $10^6$ steps each, resulting in Gaussian distributions for the fitted parameters, after ignoring the initial 20\% of each chain. We took the distributions median to be the best-fit value and the values at 84.13 and 15.87 percentiles to be the +\sig{1} and -\sig{1} confidence limits, respectively. The assumed priors and fitted parameters are listed in \tabr{fitparams}. We also applied a prayer bead analysis to the \ik\ occultation data that resulted in the same fitted parameters uncertainties as the MCMC approach.

\begin{figure}
\begin{center}
\includegraphics[scale=0.37]{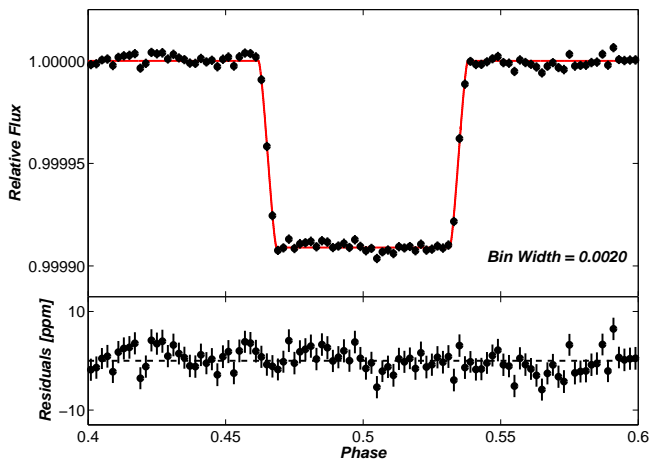}
\caption{\label{fig:occkep} 
\ik\ occultation light curve. {\it Top:} Binned short cadence data (black filled circles with error bars) and best fit model (red solid line). {\it Bottom:} Binned residuals (black circles) with error bars. A dashed black line is plotted at zero residuals, for reference.}
\end{center}
\end{figure}

The \ik\ phase folded and binned light curve is presented in \figr{occkep} along with the fitted model and residuals. The residuals scatter is 109 ppm, and since the data is 1 min cadence this scatter is also the PNR of the {\it unfolded and unbinned} data  (see \tabr{res}). The PNR of the {\it phase-folded and binned} data, equivalent to the scatter of a binned light curve using 1 min wide bins, is 5.0 ppm. We used the orbital ephemeris ($P$ and \tocc) obtained here in modeling all other data sets analyzed in this work, as it is of superior precision to any ephemeris currently available in the literature.

\begin{deluxetable}{lc}
\tablecaption{\label{tab:fitparams} Occultation fit parameters} 
\tablewidth{0pt}
\tablehead{ 
\colhead{Parameter} &\colhead{Value} \\
}
\startdata
Constrained parameters: & \\
$R_p/R_s$   & 0.0845 $\pm$ 0.0012 \\
$a/R_s$        & 4.44 $\pm$ 0.16\\
$b$                & 0.317 $\pm$ 0.033 	\\
\hline
Fitted parameters: &  \\
\dkep, ppm & 90.81 $\pm$ 0.27 \\
$P$, day         &   $ \ \, 1.76358799 \pm 0.00000037 $  \\
\tocc, BJD  &    2,455,603.448101 $ \pm$ 0.000079 \ \ \  \ \ \ \ \ \ \\
\enddata
\end{deluxetable}

%===============================================================
\subsubsection{Comparing occultation to transit times}
\label{sec:timecomp}
%===============================================================

Model fitting the transit light curve is beyond the scope of this work since it requires carefully accounting for the gravity darkening that dominates the star's surface brightness. This results from the star's rapid rotation, which is typical of main sequence star's of this spectral type. The rapid rotation leads to increased gravity near the stellar poles compared to the stellar equator, which results in increased surface temperature and hence increased brightness \citep{monnier07}. This makes the light curve deviate from the \cite{mandel02} model that includes only limb darkening, as studied in detail by \cite{barnes09} and identified for \kepb\ by \cite{szabo11}. In addition, due to the host star's rapid rotation the transit light curve is expected to be distorted also by the Photometric RM effect \citep{shporer12, groot12}. Fortunately, a detailed model fitting of \kepb\ transit was already done by \cite{barnes11}.

Comparing the mid-occultation time derived here to the mid-transit time of \cite{barnes11} shows the difference is consistent with half an orbital period:
\begin{equation}
\label{eq:deltat}
\Delta t = \tocc - \ttr - P/2 = -2.6 \pm 7.5\ {\rm s} , 
\end{equation}34
where \ttr\ is the mid-transit time from \cite{barnes11} corrected for the \ik\ timing error\footnote{\url{http://archive.stsci.edu/kepler/timing\_error.html}}. We have subtracted from the above result an integer number of orbital periods. Using \ttr\ from \cite{batalha13} gives only a slightly different value, of $\Delta t = +3.0 \pm 6.9$ s, within \sig{1} from the above, although the latter value is affected by fitting a symmetric transit light curve model to an asymmetric one, so it does not account for the effects mentioned above which could bias the mid-transit time measurement. We verified that the period derived here is consistent with the period reported in the literature based on analysis of the transit light curve \citep{borucki11, barnes11, batalha13}, although the period reported here is more precise.

The result presented in \eqr{deltat} is surprising since we would expect a time difference of $+34.0 \pm 0.7$ s due to light travel time delay \citep[e.g.,][]{loeb05, kaplan10}. This expectation is based on the system parameters measured here. The difference between the expected and measured $\Delta \rm t$  equals $(+2.4 \pm 0.5)\times10^{-4}$ of the orbital phase, close to a \sig{5} significance. 

\begin{figure}
\begin{center}
\includegraphics[width=8.5cm]{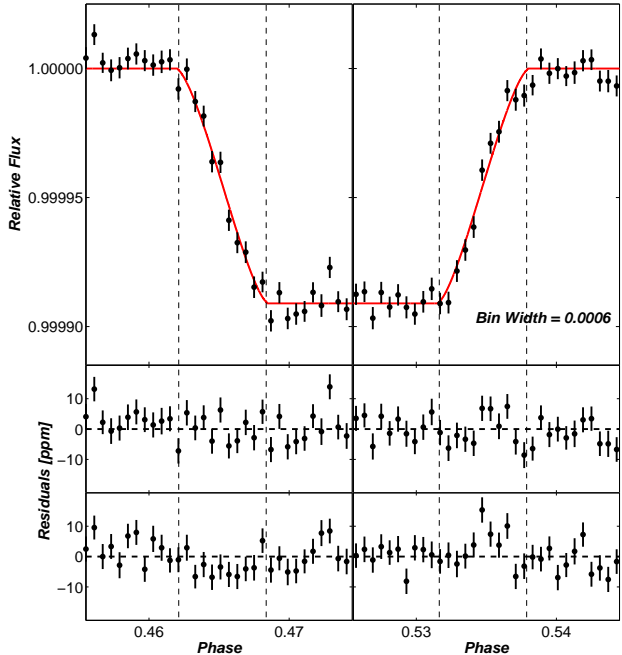}
\caption{\label{fig:occkepzoom} 
{\it Top:} Zoomed-in view of \ik\ phase-folded and binned light curve, as relative flux vs.~phase, around ingress (left) and egress (right). Red solid line marks the fitted model. {\it Middle:} Residuals, data subtracted by fitted model, of the light curves shown at the top panels. Horizontal dashed black line marks the zero residuals level, for reference. {\it Bottom:} residuals from a model shifted to the expected mid-occultation time assuming a circular orbit (see \secr{timecomp}). In all panels vertical dashed black lines mark the start and end of ingress and egress (four points of contact).
}
\end{center}
\end{figure}

To investigate this further \figr{occkepzoom} shows a zoom-in view of the \ik\ occultation light curve ingress and egress (top panels). Compared to other phases, the light curve does not show an increased scatter or correlated noise features during ingress or egress (middle panels). We also plot, in \figr{occkepzoom} bottom panels, the residuals from a model with \tocc\ shifted to the expected time and assuming a circular orbit. Those residuals do show slightly increased correlated noise features during ingress and egress, although their significance is low. This shows visually how the mid-occultation time fitted here better describes the data than the predicted mid-occultation time based on the mid-transit time of \cite{barnes11} and the light travel time delay.

One possible explanation for the measured time shift is that the light curve ingress and egress are distorted in a way that given the quality of our data it is consistent with our light curve model with a shifted mid-occultation time. Such a distortion can be induced by an asymmetric planetary optical surface brightness, where the brightest region is shifted away from the substellar point. In such cases the largest slope during ingress and egress will occur slightly earlier or later compared to cases where the planetary surface brightness is symmetric. This in turn will cause a shift in the measured mid-occultation time when using a model that assumes symmetric surface brightness. Williams et al.~(2006; see also \citealt{dewit12}) discuss the impact of a non-uniform surface brightness on the shape of the ingress and egress and the measured mid-occultation time when fitted with a model assuming a uniform distribution. A time shift induced by a so-called hot spot, or an offset of the hottest region on the planetary surface from the substellar point, was measured by \cite{agol10} at 8~\mic\ for HD~189733b. In that case the hot spot is attributed to super-rotating winds near the planetary equator that shift the hot spot eastward of the substellar point \citep[e.g.,][]{showman02, knutson07, showman09}. This causes a delay in the measured \tocc\ whereas we find that \kepb's occultation occurs early. The latter would be consistent with a bright region located westward of the substellar point. \cite{demory13} identified a non-uniform reflectivity in the optical for Kepler-7b, where the planet's most reflective region is located westward of the substellar point. This should cause the measured occultation time for that planet to occur early, although this effect is not currently detectable in the Kepler-7 system. 

A different possible explanation for the time shift is a small orbital eccentricity, $e$. A small eccentricity adds approximately $2Pe\cos{\omega}/\pi$ to the time between occultation and transit \cite[][Eq. 33]{winn11}, where $\omega$ is the argument of periastron. Therefore a small eccentricity of about only $5\times10^{-4}$ is enough to shift $\Delta t$ by about half a minute, while inducing a very small difference between the transit and occultation duration \citep[][Eq. 34]{winn11}, of less than 10~s. The impact of such a small eccentricity on the phased light curve shape (\secr{kepler_phase}) is undetectable with our current data, so we cannot reject this possibility out of hand.

Yet another process that may affect the measured \tocc\ is the propagation-delay effect described by \cite{loeb05}, in which the planet is moving away from the observer during occultation ingress and towards the observer during egress, causing the latter to appear slightly shorter than the former. However, that effect is below the sensitivity of our data.

%===============================================================
\subsection{\ik\ Phase Curve Data}
\label{sec:kepler_phase}
%===============================================================

\ik's high quality data allows us to study optical photometric modulations induced by orbital motion in star-planet  systems \citep{loeb03, zucker07}. For the \kepA\ system that was already done by several authors \citep{shporer11, mazeh12, mislis12, esteves13, placek13}. We carry out here an analysis of the optical phase curve using 13 quarters of long cadence data, Q2 through Q14, comprising over four times the \ik\ data used in previous studies. We do not use \ik\ short cadence data here since the higher time resolution has no additional value for the study of the sinusoidal variability along the orbital motion, using it will make the analysis unnecessarily more CPU-intense due to the increased amount of data points, and short cadence data is available only for 10 quarters leading to Q14.

We first remove instrumental signals, or trends, by fitting the first four cotrending basis vectors (CBVs) to the data of each quarter using the Pyke Python package \citep{still12}. We then continue in a similar way to \cite{shporer11}, where each continuous segment of data was detrended by fitting a 5th degree polynomial while ignoring in-eclipse (in-transit and in-occultation) data, and then divided by that polynomial. This did not affect the sinusoidal modulations along the orbit since the duration of each continuous segment is at least an order of magnitude longer than the orbital period. Fitting was done while iteratively rejecting \sig{5} outliers until none are identified. Using polynomial degrees of 4 and 6 did not change the results. We used the ephemeris obtained in the \ik\ occultation light curve fit (See \secr{kepler_occ} and \tabr{fitparams}), and an eclipse duration an hour longer than the known 3.2 h duration, to remove any long cadence measurements partially in-eclipse.

Next we analyze the variability in the phase folded light curve. There are three well known mechanisms through which the orbital motion of the star-planet system induces photometric modulations \citep[e.g.,][]{loeb03, zucker07, faigler11}. Those include (1) the reflection effect, due to both planetary thermal emission and reflected stellar light from the planetary surface, (2) the beaming effect, due to the varying RV of the stellar host, and (3) the ellipsoidal effect, due to tidal forces induced by the planet on the host.  

We modeled the photometric modulations along the orbit using a simple model consisting of a sinusoidal component at the orbital period and two additional sinusoidal components at the first and second harmonics:
\begin{eqnarray}
\label{eq:beer}
f(t) = a_0 + a_{1c}\cos\left(\frac{2\pi}{P}t\right) + a_{1s}\sin\left(\frac{2\pi}{P}t\right) \nonumber \\ 
+ a_{2c}\cos\left(\frac{2\pi}{P/2}t\right) + a_{2s}\sin\left(\frac{2\pi}{P/2}t\right) \nonumber \\  
+ a_{3c}\cos\left(\frac{2\pi}{P/3}t\right) + a_{3s}\sin\left(\frac{2\pi}{P/3}t\right),
\end{eqnarray}
where $f$ is relative flux, $P$ the orbital period, and $t$ is time subtracted by mid-transit time, so the transit is taken to be at orbital phase zero. This model is similar to the BEER model \citep{faigler11} which we used in \citet{shporer11}, only with an additional component at the 2nd harmonic. In this formalism the reflection, beaming, and ellipsoidal effects have amplitudes $a_{1c}$, $a_{1s}$, and $a_{2c}$, respectively. The coefficient of the sine component of the 1st harmonic, $a_{2s}$, and the coefficients of the 2nd harmonic, $a_{3c}$ and $a_{3s}$, are not associated with any of the well known physical effects and therefore expected to be small if not negligible, and are included in the model for completeness. We refer to this model as the 3-harmonics model. We also carried out a separate analysis without the 2nd harmonic, meaning where $a_{3c}$ and $a_{3s}$ are fixed to zero, and refer to that model as the 2-harmonics model. 

We phase folded the data using the occultation ephemeris derived here, and iteratively fitted the model above while rejecting outliers until no \sig{4.5} outliers were left. The outlier threshold was chosen by visually examining the phased data, consisting of approximately 44,000 individual long cadence data points, to verify that all rejected points are clear outliers and no additional outliers are left. 

\begin{figure}
\includegraphics[width=93mm]{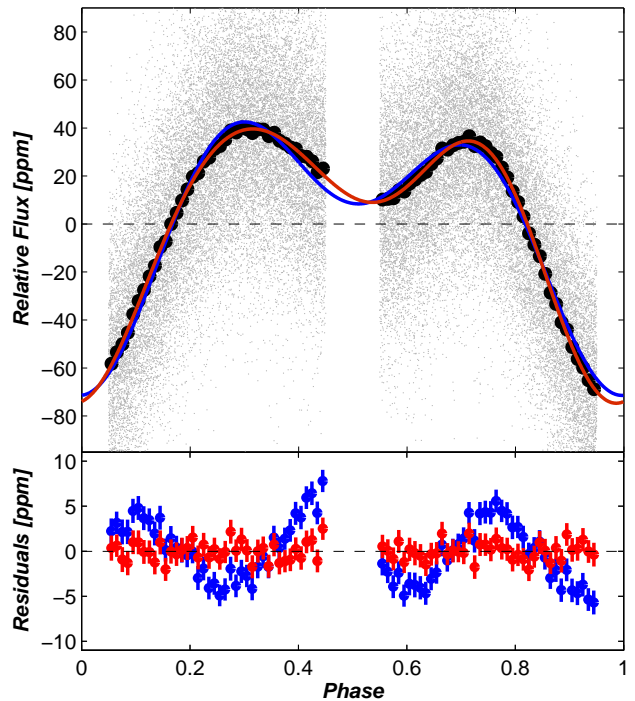}
\caption{\label{fig:orbphot} 
{\it Top:} Phase folded long cadence \ik\ light curve. Long cadence is marked by gray dots in the back ground, binned light curve is in black filled circles (error bars comparable to marker size). The blue and red solid lines are the 2-harmonics and 3-harmonics model fits, respectively. {\it Bottom:} Residuals (data subtracted by the model) from the 2-harmonics model in blue and the 3-harmonics model in red. The 2-harmonics model residuals show strong correlated noise features which are not seen in the 3-harmonics model residuals, indicating the latter is a better description of the data.}
\end{figure}

Our fitted 3-harmonics model is shown in red in \figr{orbphot}, including residuals in the bottom panel, and the fitted amplitudes listed in \tabr{amps} bottom part. The fitted 2-harmonics model is shown in blue in \figr{orbphot} and the fitted coefficients listed in \tabr{amps} upper part. The fitted values were derived using a linear least squares method, while the error bars were determined using the prayer bead approach. The median of those distributions was identical to the originally fitted values. To test our results, we repeated the analysis for each quarter separately, and then took the average between all quarters. The results of this ``quarter averaging" approach were indistinguishable from our original results.

\tabr{amps} lists the measured amplitudes in the middle column, and also the corrected amplitudes, after multiplying by the \ik\ dilution factor (see \secr{spec} and \tabr{res}), in the right-most column. The errors on the latter account for the errors in both the measured amplitudes and the dilution factor. Comparing the fitted amplitudes between the two models shows that adding an additional harmonic component in the 3-harmonics model does not change the amplitudes significantly, albeit perhaps the Reflection amplitude where the corrected amplitude increased by \sig{1.6}, which are 2.2\%.

\figr{orbphot} bottom panel shows that the 2-harmonics model residuals have significant systematic features which do not appear in the 3-harmonics model residuals, indicating the latter is a more complete model than the former. Adding additional higher harmonics to the model does not result in statistically significant fitted amplitudes. Although $a_{3c}$ and $a_{3s}$ are small, at the several ppm level, they are statistically significant. This signal, at one third the orbital period, was already identified by \citet{esteves13} with a consistent amplitude\footnote{We corrected the amplitude reported by \citet{esteves13} to account for the weaker dilution assumed by those authors.}. It is at least partially due to a higher order component of the tidal ellipsoidal distortion \citep{morris85, morris93}, although the theoretically predicted amplitude is about half the overall amplitude of the measured signal at one third the orbital period and has a different phase, so there is an additional process, or processes, in play here. This could be related to incomplete understanding of tidal ellipsoidal modulations of massive stars with radiative envelopes, especially when the stellar spin axis is not aligned with the orbital angular momentum axis and the stellar rotation rate is not synchronized with the orbit \citep{pfahl08, vankerkwijk10, jackson12}, as is the case for \kepA. Another possibility is that this signal originates from a more complicated planetary surface distribution pattern than assumed here \citep{cowan13}. A detailed investigation of the combination of these effects is beyond the scope of this study.

\begin{deluxetable}{lcc}
\tablecaption{\label{tab:amps} Fitted amplitudes} 
\tablewidth{0pt}
\tablehead{ \colhead{Component} &\colhead{Measured amplitude} & \colhead{Corrected amplitude}  \\
\colhead{} &\colhead{ppm} & \colhead{ppm} }
\startdata
\textcolor{blue}{ 2-harmonics model:} & & \\
$a_{1c}$ -- Reflection  &   $ 40.00 \pm 0.37 $      & $76.52 \pm 1.04$ \\
$a_{1s}$ -- Beaming    &   $ \ \, 4.97 \pm 0.30 $   & $\ \, 9.51 \pm 0.58$ \\   
$a_{2c}$ -- Ellipsoidal &   $ 31.40 \pm 0.48 $      & $60.07 \pm 1.10$ \\
$a_{2s}$  &   $ \ \, 0.34 \pm 0.36 $   & $\ \, 0.65 \pm 0.69$ \\
\hline 
\textcolor{red}{3-harmonics model:} & & \\
$a_{1c}$ -- Reflection  &   $ 40.87 \pm 0.36 $      & $78.18 \pm 1.04$ \\
$a_{1s}$ -- Beaming    &   $ \ \, 4.94 \pm 0.25 $   & $\ \, 9.45 \pm 0.49$ \\   
$a_{2c}$ -- Ellipsoidal &   $ 31.41 \pm 0.44 $      & $60.09 \pm 1.03$ \\
$a_{2s}$  &   $ \ \, 0.42 \pm 0.32 $   & $\ \, 0.80 \pm 0.61$ \\
$a_{3c}$  &   $ \ \, 1.66 \pm 0.29 $   & $\ \, 3.18 \pm 0.56$ \\
$a_{3s}$  &   $ \ \, 4.12 \pm 0.20 $   & $\ \, \ 7.88 \pm 0.39$ 
\enddata
\end{deluxetable}

The parameters we fitted in \cite{shporer11} are $\leq$~\sig{2} of the values derived here. Differences could arise from the changes in our analysis method. Here we used the highly precise occultation ephemeris, while in \cite{shporer11} we used an ephemeris derived from analysis of the photometric orbital modulations themselves. In \cite{shporer11} we also did not remove instrumental trends by fitting cotrending basis vectors.

%===============================================================
\subsubsection{Planetary mass estimate}
\label{sec:planetmass}
%===============================================================

The beaming and ellipsoidal phase modulation amplitudes depend linearly on the planetary mass so the latter can be estimated using the corrected amplitudes (\tabr{amps} rightmost column), $A_{\rm beam}$ and $A_{\rm ellip}$, along with other parameters of the system and the host star:
\begin{align}
\label{eq:abeam}
& \hspace{-8mm} M_{ \rm p,\ \! \!  beam}\sin i  =  & \nonumber \\
&  \hspace{2mm} \frac{0.37}{\alpha_{\rm beam}} \left( \frac{M_{ \rm s}}{M_{\sun}}\right)^{2/3} \left( \frac{P_{\rm orb}}{\rm day} \right)^{1/3} \left(\frac{A_{\rm beam}}{\rm ppm}\right) \ M_{ \rm J} ,  \\
\label{eq:aellip}
& \hspace{-8mm} M_{ \rm p,\ \! \!  ellip}\sin i = & \nonumber \\
& \hspace{-6mm} \frac{0.077}{\alpha_{\rm ellip} \sin i}  \left(\frac{R_{ \rm s}}{R_{\sun}}\right)^{-3} \left(\frac{M_{ \rm s}}{M_{\sun}}\right)^{2} \left(\frac{P_{\rm orb}}{\rm day}\right)^{2} \left(\frac{A_{\rm ellip}}{\rm ppm}\right) \ M_{ \rm J} ,  
\end{align}
where $M_{\rm s}$ and $R_{\rm s}$ are the host mass and radius, $P$ and $i$ are the orbital period and orbital inclination angle, and $\alpha_{\rm beam}$ and $\alpha_{\rm ellip}$ are order of unity coefficients. We note that $\sin i$ can be ignored since as measured by \cite{barnes11} it is close to unity, as expected for this transiting system. 

The beaming coefficient, which accounts for the photons being Doppler shifted in and out of the observed bandwidth along the orbital motion of the star, is calculated as:
\begin{equation}
\label{eq:alphabeam}
\alpha_{\rm beam} = \frac{x e^{x}}{e^{x}-1} = 0.80 \pm 0.02, \ \ \ \ \  \left(x \equiv \frac{h\nu}{k_{\rm B}\teff}\right) , 
\end{equation}
where $h$ is Planck's Constant, $\nu$ the observed frequency, $k_B$ Boltzman Constant, and \teff\ the host star's effective temperature. For the latter we used the value derived here while integrating over the \ik\ transmission curve\footnote{\url{http://keplergo.arc.nasa.gov/CalibrationResponse.shtml}}. The error reported above accounts for the error in \teff. 

The ellipsoidal coefficient is approximated as \citep{morris93}: 
\begin{equation}
\label{eq:alphaellip}
\alpha_{\rm ellip} = 0.15 \frac{(15+u)(1+g)}{3-u}  = 1.43 \pm 0.14 ,
\end{equation}
where $u$ is the stellar limb darkening coefficient assuming a linear limb darkening law, and $g$ the gravity darkening coefficient. We estimated those coefficients using the grids of \cite{claret11} and stellar parameters derived here.

\eqsr{abeam}{aellip} assume the companion is a non-luminous object, so the observed beaming and ellipsoidal modulations are entirely due to the host star's motion, not the planetary companion. This assumption is not entirely correct, as the planet does have a measurable contribution to the total flux in the optical. This contribution is at the 10$^{-4}$ level (\secr{kepler_occ}), and less during transit phase (\secr{energybudget}), so its contribution to the measured amplitudes is less than the amplitude errors even when accounting for the planetary companion's larger RV amplitude \citep{zucker07, shporer10}.

Our derived planetary masses using the beaming and ellipsoidal modulations amplitude are listed in \tabr{planetmass} where the errors account for the errors on all parameters in the right hand side of \eqsr{abeam}{aellip}, leading to significantly larger fractional errors for the mass estimates than for the beaming and ellipsoidal corrected amplitudes (see \tabr{amps}). The two estimates are \sig{1.8} from each other\footnote{The statistical significance of the difference between the two mass estimates was calculated while accounting for the fact that they are \emph{not} independent. For example, they are both derived using the same stellar mass and dilution factor.} and differ by a factor of $1.27\pm0.23$. Similar discrepancies were reported by other authors for other systems \citep[e.g.,][]{vankerkwijk10, carter11, bloemen12, barclay12, faigler13, esteves13}, and for this system by \cite{mazeh12} using only two quarters of \ik\ long cadence data. The reason for this discrepancy is currently not clear. One possibility is inaccurate stellar parameters, meaning a poor understanding of the nature of the host star which is more probable for early-type stars where accurate stellar parameters (e.g., mass, gravity, temperature) are difficult to obtain compared to Sun-like stars. The refined stellar parameters and dilution factor obtained here result in a smaller beaming-based planet mass estimate compared to our results in \cite{shporer11}. 

Another possibility for the origin of the discrepancy is a poor understanding of the ellipsoidal effect for hot stars such as \kepA\ \citep{pfahl08, vankerkwijk10}. Further reason to doubt the ellipsoidal mass estimate comes from \cite{ehrenreich11} and \cite{bloemen12}, who measured the RV amplitude of the A-type primary in KOI-74, and showed it agreed with the beaming amplitude and did not agree with the ellipsoidal amplitude \citep{vankerkwijk10, bloemen12}. However, one should be careful when comparing \kep\ and KOI-74. The A-type primary in KOI-74 is almost 2,000 K hotter than \kepA\ and the secondary is not a planet but a white dwarf \citep{vankerkwijk10, ehrenreich11, bloemen12}, leading to observed photometric amplitudes a few orders of magnitude larger. On the other hand, some authors have reported cases where the known RV amplitudes were consistent with the ellipsoidal-based mass estimate and inconsistent with the beaming-based mass estimate \citep[e.g.,][]{faigler13, esteves13}. Although all the latter cases involve convective Sun-like stars. 

We therefore take a conservative approach and conclude that our planet mass estimate lies within the range of 4.94 -- 8.09 \mjup\ with a \sig{1} confidence. This estimate, derived using the refined stellar parameters and dilution factor obtained here is consistent with previous estimates \citep{shporer11, mazeh12, mislis12, esteves13, placek13}, confirming this planet belongs to the rare class of massive hot Jupiters. The A-type nature of the host star might represent an extension of the tendency of hot Jupiters at this mass range and above, into the brown dwarf mass range, to orbit F-type stars as opposed to G-type Sun-like convective stars \citep{bouchy11a, bouchy11b}.

\begin{deluxetable}{lc}
\tablecaption{\label{tab:planetmass} Companion mass estimates} 
\tablewidth{0pt}
\tablehead{ \colhead{Method} &\colhead{Value} 
}
\startdata
$M_{ \rm p,\ \! \!  beam}\sin i$, \mjup     &   $ 7.57 \pm 0.52 $  \\
$M_{ \rm p,\ \! \!  ellip}\sin i$, \mjup        &   $ 5.94 \pm 1.00 $ \\
\enddata
\end{deluxetable}

%===============================================================
\subsection{Spectroscopic Analysis}
\label{sec:spec}
%===============================================================

As we have already noted, since the two A-type stars are at  a separation of 1\farcs15$\pm$0\farcs05 (see \figr{fchart} and \secr{kepler13}) they are fully blended in all our photometric data sets. A correct astrophysical interpretation of any measured photometric variability requires an estimate of the ``real" variability amplitude that would have been measured if the star was resolved. Obtaining those amplitudes from the directly measured ones requires knowledge of the flux ratio between the two stars in the observed bands. We estimate the magnitude of this dilution using two high-resolution spectra of the two stars, taken with the Keck/HIRES spectrograph \citep{vogt94} on UT June 3, 2010. The spectra were taken while placing the slit perpendicular to the position angle between the two stars. Given the slit width of 0\farcs86 and the typical 0\farcs5 astronomical seeing at the Keck Observatory we estimate the contamination between the two spectra is no more than 10\%, and we did not see any signs of contamination in our visual examination of the raw spectra. 

According to \cite{santerne12} the third star in the system, \kepBB, contributes very little to the overall system luminosity, from approximately 1\% in the optical to 2\% in the IR. We therefore ignore its existence in the following analysis and treat the system as consisting of two A-type stars.

The spectra of components A and B were fitted individually to derive precise stellar parameters, and to predict the flux ratio between the two components throughout a wide range of wavelengths, from the optical to the IR. In this step, we made use of the Phoenix spectral models \citep{hauschildt99, husser13}, and followed the fitting recipe described in \cite{szabo11}. Unfortunately the rapid rotation of both stars meant that the metal lines were blended, resulting in degeneracies in our retrieved stellar parameters. The temperatures of the two components were determined to be 7,650 K and 7,530 K, with 250~K errors for both values. Although the $\chi^2$ surface of the fit in the \teff\ -- \logg\ grid was wide with a shallow bottom, the error surfaces could have been shifted to each other with a temperature shift of $120\pm50$~K. This means that the spectra better constrained the temperature difference than the individual temperatures. A temperature difference of $120\pm50$~K is also consistent with the results of \cite{szabo11} from optical photometric colors.

To improve the fits, we assumed that the two stars have identical ages and metallicities, and used the average density of \kepA\ determined from the transit light curve \citep[e.g.,][]{winn11}. Since the geometry of \kepb\ orbit is known and the transit chord is precisely determined \citep{barnes11}, the transit duration constrains the density of the host star. Following the recipe of \cite{winn11} the derived density of \kepA\ is 0.53$\pm$0.01~g~cm$^{-3}$, 0.37 times the solar value. The 2\% error is mostly due to the uncertainty in the impact parameter.

\begin{figure}
\begin{center}
\includegraphics[scale=0.65, angle=270]{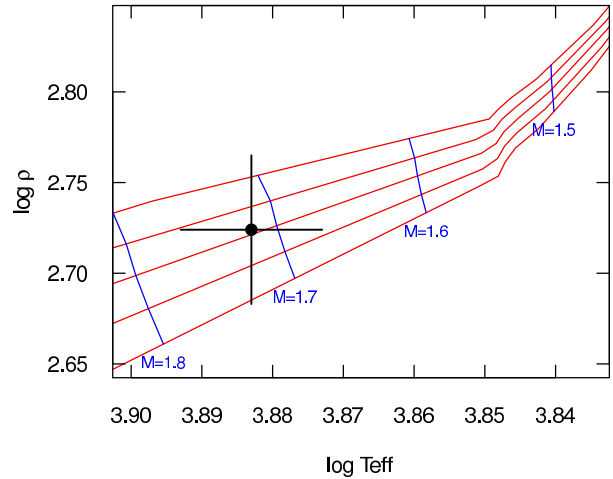}
\caption{\label{fig:rho_teff} 
Position of \kepA\ (black circle and \sig{3} errorbars) in the density (in kg m$^{-3}$) vs.~effective temperature diagram. Red  lines are Padova isochrones in steps of 100 Myr, from 300 Myr at the top to 700 Myr at the bottom, and blue lines are equal stellar mass lines (see labels).
}
\end{center}
\end{figure}

Knowing the temperature and density of \kepA, we can plot its position on the appropriate density--temperature isochrones. Stellar models were taken from the Padova isochrone family with [Fe/H]=0.2 metallicity \citep{bertelli08, bertelli09}. We find a good fit to the 0.5~Gyr old isochrones, while 0.4 and 0.6 Gyr models were beyond the \sig{1} error, as shown in \figr{rho_teff} where we plot the \sig{3} errorbars. Our results therefore point to a $0.5 \pm 0.1$~Gyr age for the \kep\ system, consistent with \cite{szabo11}. Stellar parameters are listed in \tabr{stellarparams}. 

\begin{deluxetable}{lcc}
\tablecaption{\label{tab:stellarparams} Stellar Parameters} 
\tablewidth{0pt}
\tablehead{ 
\colhead{Parameter} &\colhead{\kepA} &\colhead{\kepB} \\
}
\startdata
T$_{\rm eff}$, K & 7,650$ \ \pm \ 250$ \ \ \! & 7,530$\ \pm \ 250$ \ \ \!\\
$\log(g\ \rm [g\ cm^{-2}])$ & $4.2 \pm 0.5$ &$4.2 \pm 0.5$ \\
$\rm [Fe/H]$ & $0.2 \pm 0.2$ &$0.2 \pm 0.2$ \\
$V_{\rm rot}$, km s$^{\rm -1}$ & $78\pm15$ & $69\pm13$\\
Age, Gyr & $0.5 \pm 0.1$ & $0.5 \pm 0.1$ \\
Mass, \msun & $1.72 \pm 0.10$ & $1.68 \pm 0.10$\\
Radius, \rsun & $1.71 \pm 0.04$ & $1.68 \pm 0.04$ \\
\enddata
\end{deluxetable}

% param difference:
%T_eff(A)-T_eff(B) = 120 +- 50K
%R(A)-R(B) = 0.039 +- 0.016 Ro

\begin{figure}
\begin{center}
\includegraphics[scale=0.45]{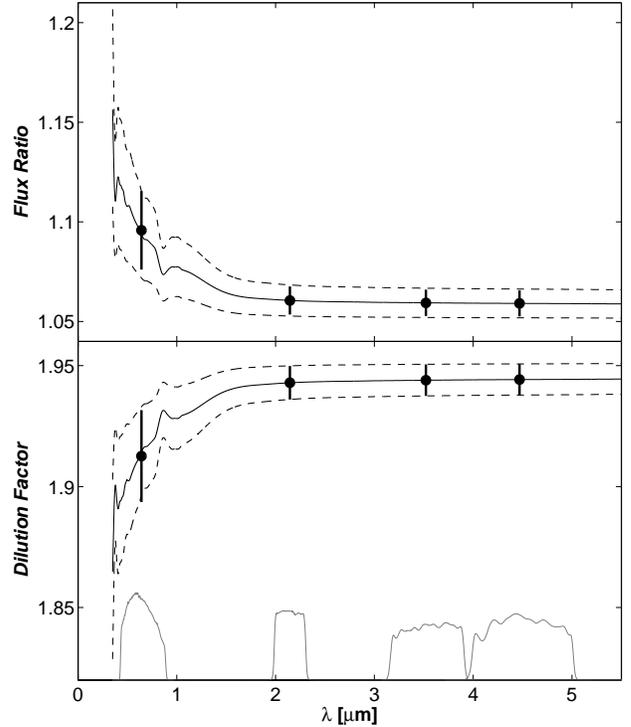}
\caption{\label{fig:dilution} 
{\it Top:} Flux ratio between the planet-hosting star \kepA\ to its binary companion \kepB, as a function of wavelength. {\it Bottom:} The multiplicative factor that corrects the measured variability amplitude (occultation depths and sinusoidal variability) to account for the dilution of the two stars, as a function of wavelength. In both panels, curves were smoothed using iterative Savizty-Golay smoothing for presentation purposes. Dashed lines represent the \sig{1} uncertainties, and filled circles with error bars are the values integrated over the bandpasses of our observations.  These bandpasses are shown at the bottom of the figure, with an arbitrary Y axis scale for presentation purposes.
} 
\end{center}
\end{figure}

The final step was to simulate the spectra of stars with masses, radii, and temperatures as indicated in \tabr{stellarparams}, assuming $v\sin i$=65~\kms\ \citep{szabo11}. For this we used Phoenix models scaled by the areas of the stellar disks and interpolated to temperatures of 7,650~K and 7,530~K. This way the flux ratios were determined for a wide wavelength range with high resolution, plotted in \figr{dilution} and listed in \tabr{fluxratio}. We calculated the flux ratio for each band as the weighted average of the flux ratio across the relevant wavelength range using the known transmission curves as weights. \tabr{res} lists the dilution factor  --- the factor by which each occultation depth needs to be multiplied to correct for the dilution --- for each of the four bands used here.

We note that using the revised planet host star's radius derived here and the planet to star radii ratio from \cite{barnes11} we obtain a revised planet radius of $R_p = 1.406 \pm 0.038$ \rjup. We also use the revised planet host star parameters in deriving the planet's mass in \secr{planetmass}.

\begin{deluxetable}{cccc} 
\tablecaption{\label{tab:fluxratio} Kepler-13A / Kepler-13B Flux Ratio}  
\tablewidth{0pt} 
\tablehead{  
\colhead{Wavelength} & \colhead{Flux ratio} & \colhead{Upper limit} & \colhead{Lower limit} \\ 
\colhead{\AA} & \colhead{} & \colhead{} & \colhead{} \\ 
} 
\startdata 
 3500.087  &  1.148113  &  1.194621  &  1.102890  \\
 3500.239  &  1.149844  &  1.197161  &  1.103864  \\
 3500.391  &  1.150445  &  1.198042  &  1.104202  \\
 3500.543  &  1.150080  &  1.197507  &  1.103997  \\
 3500.695  &  1.149214  &  1.196236  &  1.103509  \\
 3500.848  &  1.148232  &  1.194795  &  1.102956  \\
 3501.000  &  1.147166  &  1.193231  &  1.102356  \\
 3501.152  &  1.145494  &  1.190779  &  1.101415  \\
 3501.304  &  1.143709  &  1.188160  &  1.100410  \\
 3501.457  &  1.142740  &  1.186738  &  1.099864  \\
\enddata 
\tablecomments{Columns include, from left to right: Wavelength, flux ratio, flux ratio \sig{1} upper boundary, and flux ratio \sig{1} lower boundary. The table in its entirety is available on-line.}
\end{deluxetable}

\begin{deluxetable*}{cccccc}
\tablecaption{\label{tab:res} Occultation depths and light curves' scatter. } 
\tablewidth{0pt}
\tablehead{\colhead{Band} &\colhead{Scatter} & \colhead{PNR} & \colhead{Measured depth} & \colhead{Dilution factor} & \colhead{Corrected depth}  \\
\colhead{}   &\colhead{\%} & \colhead{\% min} & \colhead{\%} & \colhead{} & \colhead{\%} }
\startdata

IRAC/4.5~\mic   &  0.339 & 0.141  &  $ 0.114 \pm 0.012 $                & $1.9442 \pm 0.0064$ & $0.222 \pm 0.023$\\ 

IRAC/3.6~\mic   &  0.442 & 0.105  &   $ 0.080 \pm 0.016 $               & $1.9440 \pm 0.0065$ & $0.156 \pm 0.031$ \\ 

WIRC/\ks  &   0.360 & 0.206 &   $ 0.063 \pm 0.026 $               & $1.9429 \pm 0.0070$ & $0.122 \pm 0.051$ \\

\ik\                &  0.0109\tablenotemark{a} & 0.00050\tablenotemark{b}  &   $ 0.009081 \pm 0.000027 $ & $1.913 \pm 0.019$  & $0.01737 \pm 0.00018$\\
\enddata
\tablenotetext{a}{Scatter (MAD) in the unbinned short cadence light curve.}
\tablenotetext{b}{PNR (photometric noise rate) of the phase folded and binned light curve, using 1 min bins.}
\end{deluxetable*}

%%%%%%%%%%%%%%%%%%%%%%%%%%%%%%%%%%%%%%%%%
\section{Atmospheric Characterization}
\label{sec:atm}
%%%%%%%%%%%%%%%%%%%%%%%%%%%%%%%%%%%%%%%%%

We studied the atmosphere using two different approaches. In the first (\secr{energybudget}) we used a highly simplistic model where the small number of parameters allowed us to fit them despite the small number of data points. In the second (\secr{atm_model}) we used models with many parameters so we could not fit the models to our data, but only do a qualitative comparison. 

%%%%%%%%%%%%%%%%%%%%%%%%%%%%%%%%%%%%%%%%%
\subsection{Energy Budget}
\label{sec:energybudget}
%%%%%%%%%%%%%%%%%%%%%%%%%%%%%%%%%%%%%%%%%

\begin{figure}
\begin{center}
\includegraphics[scale=0.46]{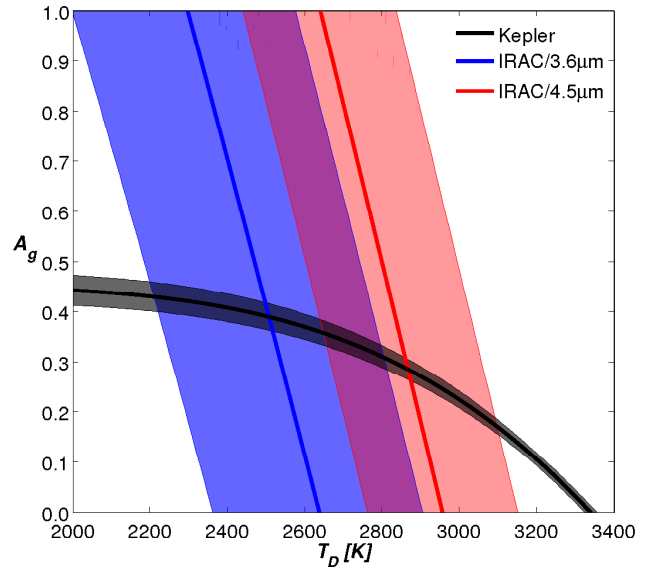}
\caption{\label{fig:ag_td} 
Geometric albedo versus day-side brightness temperature for three of the observed bands, \ik\ in black, IRAC/3.6~\mic\ in blue, and IRAC/4.5~\mic\ in red. The solid lines are calculated according to \eqr{depth} and the measured occultation depths, and the filled regions are the corresponding \sig{1} uncertainty while marginalizing over the parameters in \eqr{depth}. The area where the IRAC/3.6~\mic\ and IRAC/4.5~\mic\ \sig{1} regions overlap is marked in purple. The WIRC/\ks\ band occultation measurement is ignored here since its low S/N results in a wide \sig{1} region that encompasses both those of the IRAC/3.6~\mic\ and IRAC/4.5~\mic\ so it does not add information.
}
\end{center}
\end{figure}

The day-side luminosity is a combination of thermal emission and reflected stellar light. Therefore it depends on the brightness temperature, \td, and geometric albedo, \ag, in the corresponding wavelength. The expected occultation depth, $D$, in relative flux, is the ratio between the planet's day-side luminosity and the star's flux, and is a function of three variables:
\begin{eqnarray}
\label{eq:depth}
D(\lambda, \td, \ag) =  \left(\frac{\rpl}{\rstar}\right)^2 \frac{B_p(\lambda, \td)}{I_s(\lambda)} + \ag\left(\frac{\rpl}{a}\right)^2 , 
\end{eqnarray}
where we take the planet's day-side emission spectrum, $B_p$, to be a black body spectrum so it depends only on wavelength, $\lambda$, and \td. We use a Phoenix stellar atmosphere model for the star's spectrum ($I_s$) interpolated to match the $T_{\rm eff}$ and $\logg$ from \tabr{stellarparams}. Since the geometric parameters of the system ($R_p/R_s$, $a/R_s$) are known the expected occultation depth at a given wavelength depends on two variables, \td\ and \ag. In \figr{ag_td} we plot the relation between these two variables for each of the observed bands, including the \sig{1} region while marginalizing over all the parameters in \eqr{depth}. We exclude the WIRC/\ks\ occultation measurement in \figr{ag_td} since due to that measurement's low S/N  the corresponding \sig{1} region encompasses both those of the IRAC/3.6~\mic\ and IRAC/4.5~\mic\ regions so it does not add any information. The \ag--\td\ relation for the \ik\ band goes from two extreme scenarios. A cold atmosphere with a high geometric albedo where the day-side luminosity is dominated by reflected stellar light on one end, and a hot atmosphere with a negligible albedo where the day-side luminosity is dominated by thermal emission on the other end. The almost vertical shape of the \ag--\td\ curves for the two \is\ bands means that in the IR the day-side luminosity is dominated by thermal emission and depends only weakly on the geometric albedo.

None of the four bands in which we observed the occultation is at the peak of the black body spectrum for a body with an effective temperature similar to that of \kepb\ atmosphere\footnote{According to Wien's displacement law, for effective temperatures in the range of 2,500 K to 3,000 K the black body spectrum peaks at 1.16~\mic\ to 0.97~\mic.}. However, the four bands sample both sides of the peak: Wien's tail in the optical and the Rayleigh-Jeans tail in the IR. Therefore if is interesting to check if the same day-side brightness temperature can reproduce all four measured occultation depths since that will give an estimate of the equivalent of the day-side {\it effective} temperature, \tdeff. We do not need to assume here that the day-side atmosphere behaves as a black body. By definition \tdeff\ is the effective temperature of a black body that shows the same occultation depths in the four wide bands we measured. We do need to assume, however, that \ag\ remains the same across the observed bands since this requires simultaneously fitting \tdeff\ and \ag\ to all four occultation depths, using \eqr{depth}. Visually, the allowed values correspond to the overlap region in \figr{ag_td}. Fitting using a dense 2-dimensional grid results in $\tdeff$ = 2,750 $\pm$ 160 K and $\ag = 0.33^{+0.04}_{-0.06}$, where the two variables are highly correlated. Although formally we assumed here that \ag\ is the same across the wavelength range we observed in, in practice since in the IR \td\ depends weakly on \ag\ the above result for \tdeff\ will change only in case of a large variation in \ag\ from the optical to the IR. Assuming a brightness temperature equal to the derived \tdeff, varying \ag\ from 0.0 to 0.5 will change the IRAC/3.6\mic\ occultation depth by no more than \sig{1}, while the corresponding \ag\ range for the IRAC/4.5~\mic\ band is from 0.0 to 1.0. Meaning, \ag\ is constrained primarily by the occultation depth in the optical.

%%%%%%%%%%%%%%%%%%%%%%%%%%%%%%%%%%%%%%%%%
\subsection{Detailed Atmospheric Modeling}
\label{sec:atm_model}
%%%%%%%%%%%%%%%%%%%%%%%%%%%%%%%%%%%%%%%%%

\begin{figure*}
\begin{center}
\includegraphics[scale=0.41]{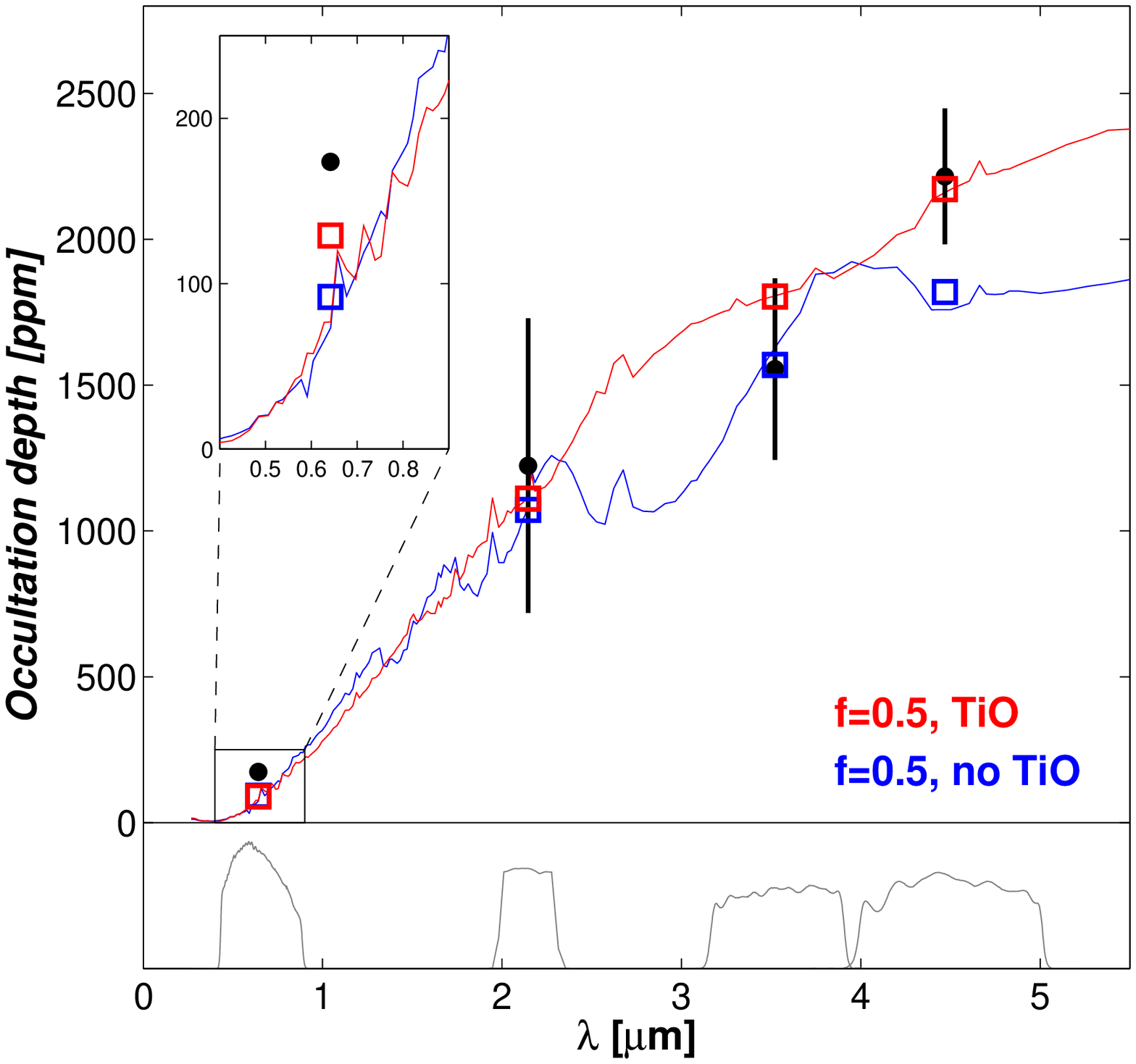}
\includegraphics[scale=0.41]{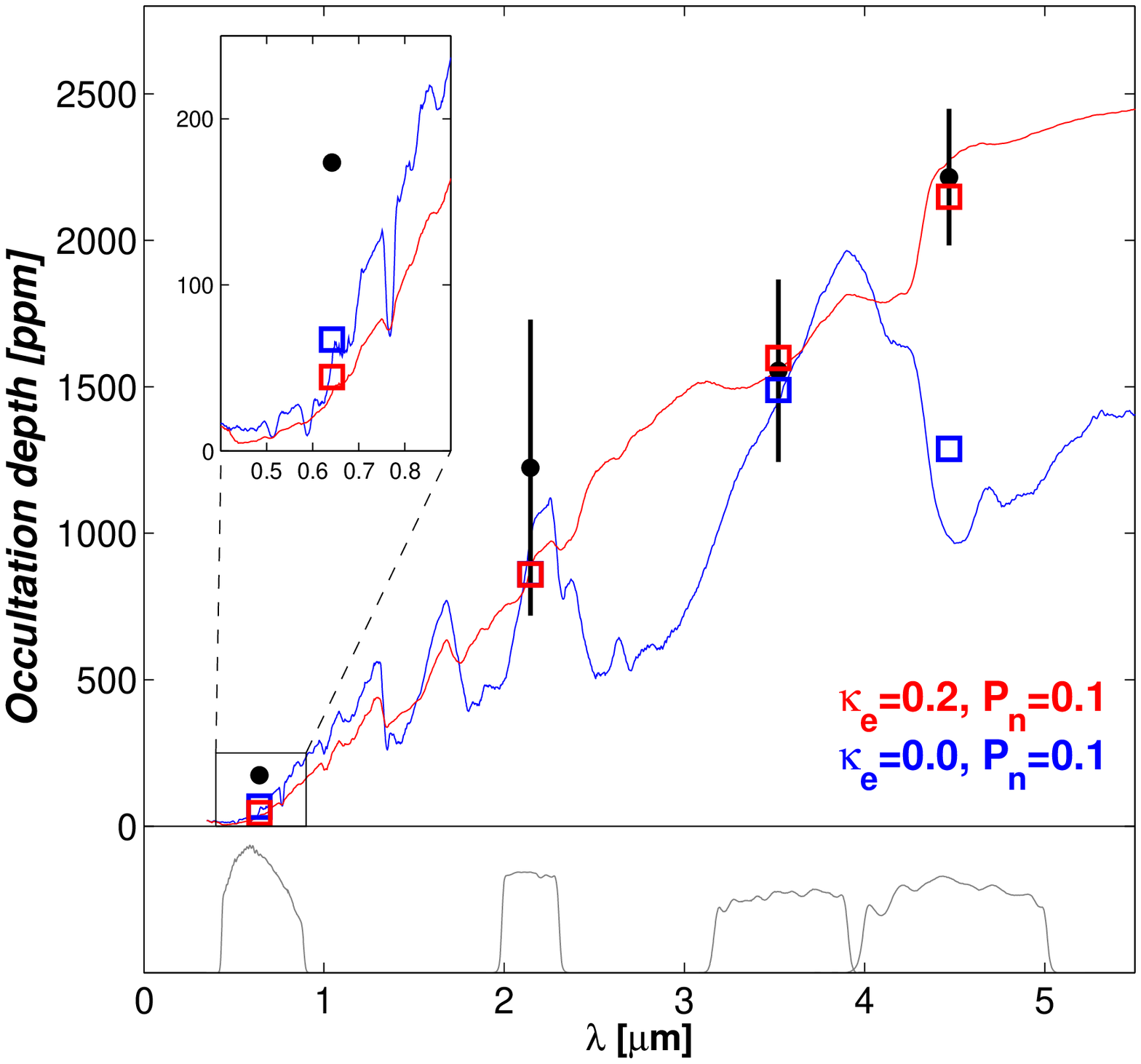}
\caption{\label{fig:atm_model} 
Occultation depth vs.~wavelength for \kepb, showing atmospheric models (solid lines, color coded) for different parameterizations (see legend). Occultation depths measured here are marked in black. Transmission curve of each band is plotted in gray at the bottom. The inset shows a zoom-in on the \ik\ wavelength region, while the error bar on \ik\ occultation depth is too small to be seen in this scale. Left panel shows models based on \cite{fortney08}, and right panels based on models of \cite{burrows08}.
}
\end{center}
\end{figure*}

In \figr{atm_model} we compare the occultation depths measured here to the expected occultation depth as a function of wavelength from detailed atmospheric models. The left panel in \figr{atm_model} shows two models based on \cite{fortney08} using a one-dimensional, plane parallel atmosphere code and assuming local thermodynamic equilibrium (LTE), and solar composition. The two models shown both assume even heat distribution across the day side and no recirculation to the night side, parametrized by \cite{fortney08} as $f$=0.5. In one model TiO is added in equilibrium abundances to the upper atmosphere, where it acts as an absorber that induces a high-altitude temperature inversion. In the second model TiO is excluded, as this species may be depleted due to cold traps on the night side and in the deep interior. We also consider models from \cite{burrows08}, which similarly assume a plane-parallel atmosphere, LTE, and solar composition, and uses opacity calculations of \cite{sharp07}. Those are shown in the right panel of \figr{atm_model}, and are parametrized by $\kappa_e$, an absorption coefficient of a gray absorber in the stratosphere, and $P_n$, an energy redistribution coefficient, ranging from 0.0 for redistribution across the day side only to 0.5 for redistribution across the day and night hemispheres. 

Comparing the two sets of models with our measured occultation depths shows that the inverted models are more consistent with the data than the non-inverted models, although that is based only on the 4.5 \mic\ point. An inverted atmosphere for \kepb\ is consistent with the trend identified by \cite{knutson10}, that planets orbiting chromospherically quiet stars tend to have inverted atmospheres. Although it is a different proxy for stellar activity, the \ik\ light curve shows an especially quiet star, and apart from variability induced by the orbiting planetary companion (\secr{kepler_phase}) the light curve shows only a $\sim$10 ppm sinusoidal variability at a period of 1.06 days \citep{shporer11, szabo11, mazeh12}. That low level variability could be due to stellar activity, but may also be due to stellar pulsations. It is also possible that it originates from the \kepB\ system as the light from all stars in the system is fully blended in the \ik\ light curve. 

The correlation identified by \cite{hartman10}, between increased planetary surface gravity and increased host-star chromospheric activity, is not supported by the \kepA\ system, as the planet has a high surface gravity and the host star shows low level variability in the light curve. Although, we note that both the \cite{knutson10} and \cite{hartman10} correlations were identified only for Sun-like convective stars, not for early-type stars like \kepA\ .

%%%%%%%%%%%%%%%%%%%%%%%%%%%%%%%%%%%%%%%%%
\subsection{Night-side Optical Luminosity}
\label{sec:nightside}
%%%%%%%%%%%%%%%%%%%%%%%%%%%%%%%%%%%%%%%%%

The \ik\ measurements of both the occultation depth and reflection modulation amplitude allows to measure the planet's night-side optical luminosity. The occultation depth measures the luminosity of the day side hemisphere, while the reflection amplitude measures the difference between the day and night hemisphere luminosities. The difference between the occultation depth and reflection (full) amplitude gives the night-side luminosity:
\begin{equation}
\label{eq:ns}
\Delta_{\rm NS} = \Delta_{\rm DS} - 2 \times { A_{\rm ref}} =  17.36 \pm  1.48 \ \ {\rm ppm} ,
\end{equation}
where $\Delta_{\rm NS}$ is the night-side luminosity and $\Delta_{\rm DS}$ the day-side luminosity. The result presented in \eqr{ns} is corrected for the dilution in the \ik\ band (see \tabr{res}). This detection of the night-side optical luminosity is beyond \sig{11} and is consistent with previous estimates \citep{mazeh12, esteves13}.

Next we derive the planet's night-side brightness temperature, \tn, using the above night-side optical luminosity, the host star's parameters (see \tabr{stellarparams}), the planet to star radii ratio (see \tabr{fitparams}), and \ik's transmission curve. We assumed a black-body spectrum for the planet and used Phoenix stellar synthetic spectra \citep{husser13} for the star, while marginalizing across the grid points close to the star's \teff\ and $\log g$, resulting in \tn\ = 2,537 $\pm$ 45 K. Using a black body spectrum for the star gives the same \tn.

While \tn\ is smaller than \tdeff\ derived above, the difference is not large. Assuming the night-side effective temperature equals the night-side brightness temperature as measured in the \ik\ band, we can use the formalism of \citet[][see their Eqs.~4~\&~5]{cowan11} to constrain the day-side to night-side heat redistribution coefficient \eps\ together with the bond albedo \ab. Using that formalism we get a heat redistribution coefficient of $\eps = 0.88 \pm 0.10$, and bond albedo \ab $\leq 0.14$ at \sig{1}. The high \eps\ indicates an efficient heat redistribution process, although the low \ab\ is inconsistent with the high \ag\ derived above assuming \ab\ = (3/2)\ag\ (Lamert's Law), which results in \ab\ = $0.50^{+0.06}_{-0.09}$. On the other hand, assuming the latter value for \ab\ requires \eps \textgreater 1 in order to reach the derived \tn. This shows that the night-side brightness temperature, measured only in the optical, is not a good proxy for the night-side effective temperature, meaning that the night side does not behave like a black body in the optical. Yet, a high \eps\ cannot be ruled out here.

%%%%%%%%%%%%%%%%%%%%%%%%%%%%%%%%%%%%%%%%%
\subsection{Comparison with WASP-33b}
\label{sec:wasp33}
%%%%%%%%%%%%%%%%%%%%%%%%%%%%%%%%%%%%%%%%%

It is interesting to compare \kepb\ to WASP-33b \citep{collier10, kovacs13}, as these are currently the only two known transiting hot Jupiters orbiting bright A-type stars. The planets in both systems orbit similar stars, have orbital periods in the 1--2 days range, have consistent (at the \sig{1} level) equilibrium temperatures, and experience consistent incident flux at the planet surface. Both systems also have orbits that are misaligned with the host stars' spin axes \citep{collier10, barnes11}, suggesting a similar orbital evolution scenario. WASP-33 may also be part of a binary star system  \citep{moya11}, although its smaller and cooler candidate companion has not been confirmed with common proper motion measurements. WASP-33b has a mass of $3.266 \pm 0.726$~\mjup\ and a radius of $1.679^{+0.019}_{-0.030}$ \rjup\ \citep{kovacs13}. Its smaller mass and slightly larger radius than \kepb\ leads to a mean density 3--4 times smaller, and a surface gravity ($\log_{10}g$) 0.3--0.5 smaller. 

Occulations of WASP-33b have been observed in the \is\ IRAC/3.6~\mic\ and IRAC/4.5~\mic\ bands \citep{deming12}, the \ks\ band \citep{deming12, demooij13}, and a narrow band S[III] filter at 0.91~\mic\ \citep{smith11}. In this case measurements of the occultation depth are hampered by the host star's $\delta$-Scuti pulsations, measured in the $R$ band by \cite{herrero11} to have a primary modulation component at 1.1~h with an amplitude of 0.1\% (see also \citealt{vonessen14}), seen also in the IR \citep{deming12}. These pulsations can bias estimates of WASP-33b occultation light curve shape, especially when based on a single occultation event which is 2.8~h long and its depth ranges from 0.4\% at 4.5~\mic\ down to 0.1\% at 0.91~\mic. 

\begin{figure*}
\begin{center}
\includegraphics[scale=0.41]{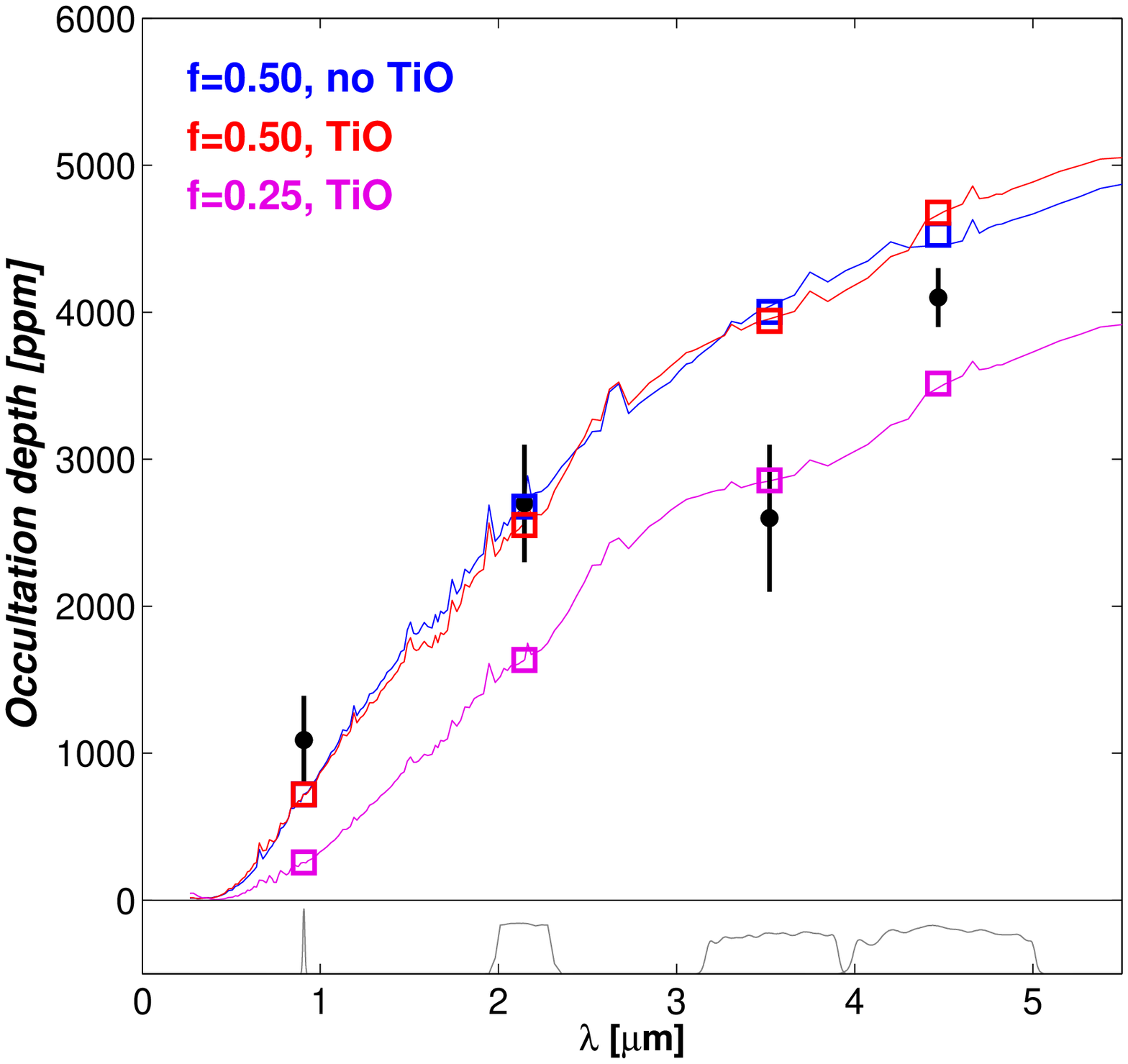}
\includegraphics[scale=0.41]{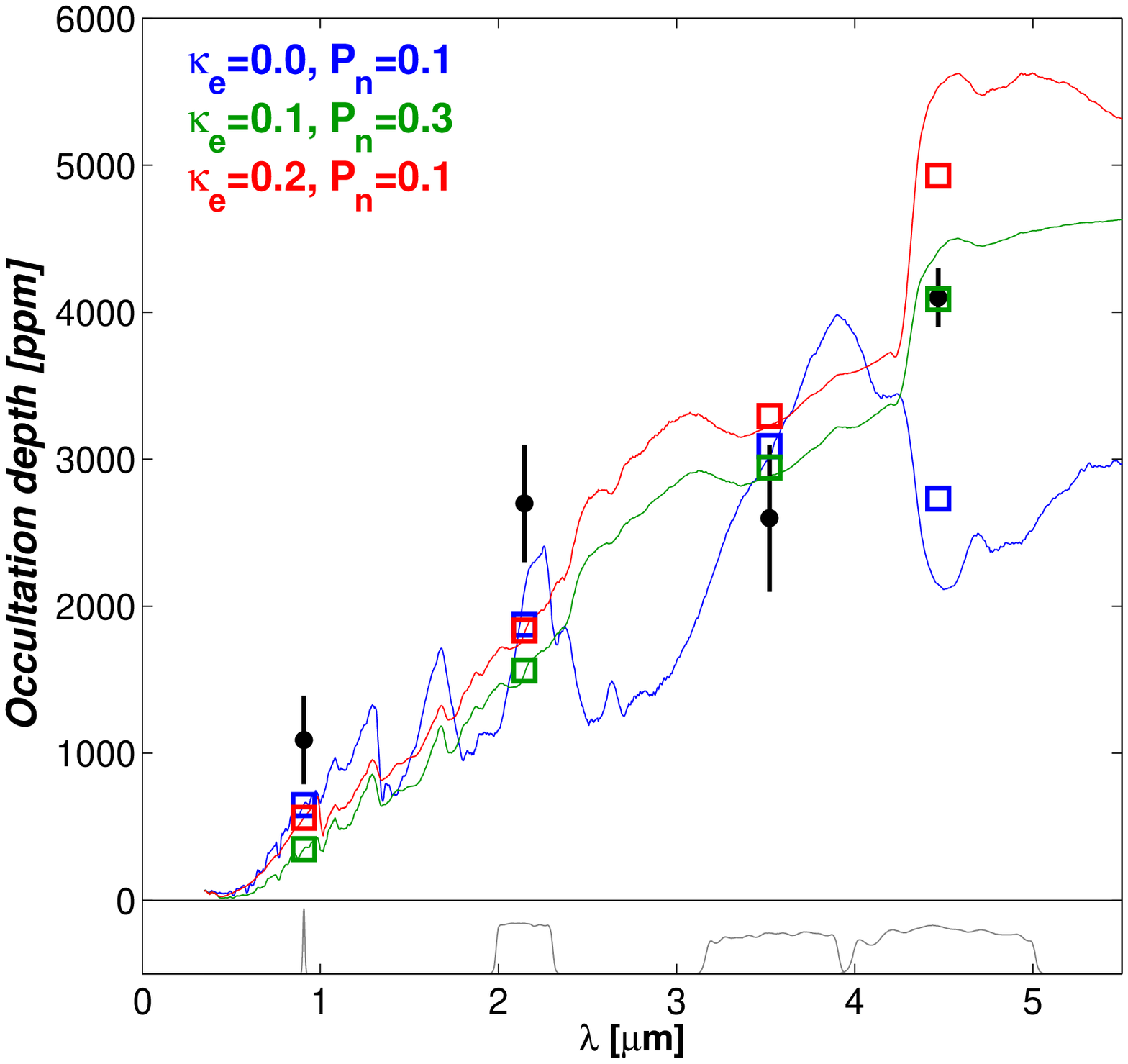}
\caption{\label{fig:wasp33} 
Occultation depth vs.~wavelength for WASP-33b, showing atmospheric models (solid lines, color coded) for different parameterizations (see legend). Occultation depths from the literature are marked in black. Transmission curve of each band is plotted in gray at the bottom. Left panel shows models based on \cite{fortney08}, and right panels based on models of \cite{burrows08}.
}
\end{center}
\end{figure*}

To compare the atmospheres of the two planets we have calculated the WASP-33b atmospheric models corresponding to those plotted in \figr{atm_model} for \kepb, and plotted them in \figr{wasp33} along with additional models with different parameters. The figure shows that none of the models are consistent with all four measured occultation depths, although the data appear to be closest to models with moderate to weak day-night recirculation and a modest temperature inversion. In this case it is possible that the discrepancies between the models and the measured occultation depths are the result of biases introduced by the $\delta$-Scuti pulsations of the host star; our group will test this in the near future with full-orbit \is\ phase curve observations including two occultations in both the IRAC/3.6~\mic\ and IRAC/4.5~\mic\ bands. \cite{deming12} explored fitting the WASP-33b occultation depths with the models of \cite{madhu09, madhu10}, which allow the relative abundances of water, methane, CO, and CO$_2$ as well as the vertical pressure-temperature profile to vary as free parameters in the fit.  They find that their models generally prefer inefficient day-night recirculation, and in the case of solar metallicity composition they prefer a day-side temperature inversion.  They also identify a family of models that match the data without a temperature inversion and with a super-solar C to O ratio, shown by \cite{madhu12} to fit the data marginally better.

%%%%%%%%%%%%%%%%%%%%%%%%%%%%%%%%%%%%%%%%%
\section{Discussion and Conclusions}
\label{sec:dis}
%%%%%%%%%%%%%%%%%%%%%%%%%%%%%%%%%%%%%%%%%

We present here a multi band study of the atmosphere of \kepb. We measured the occultation depth of \kepb\ in four wide bands, from the IR (\is/IRAC 4.5~\mic\ and 3.6~\mic), through the NIR (P200/WIRC/\ks), to the optical (\ik). We also used \ik\ data along the entire orbit to measure the planetary reflected and thermally emitted light, and measure the planetary mass from the beaming and ellipsoidal effects. Finally, we used Keck/HIRES spectroscopic data to characterize the planet host star and calculate the dilution of the observed occultations due to the presence of a blended A-type companion star. The measured occultation depths are listed in \tabr{res} including both the directly measured depths, not accounting for the dilution, and the corrected depths while accounting for the dilution. The corrected depths are simply the measured depths multiplied by the dilution factor at the respective wavelength, also listed in \tabr{res}. The corrected depth uncertainties account for both the measured depth uncertainties and dilution factor uncertainties.

Comparing the \ik\ mid-occultation time derived here (see \tabr{fitparams} and \secr{kepler_occ}) to the mid-transit time from the literature shows that the mid-occultation time occurs about half a minute earlier, with a significance of almost \sig{5} (\secr{timecomp}). This can be attributed to either a non-zero orbital eccentricity, or, an asymmetric distortion in the light curve ingress and egress shape due to asymmetric planetary surface brightness distribution, meaning the brightest region on the planet's surface is shifted away from the substellar point. An early mid-occultation time can be the result of super-rotating winds causing the planet's most reflective region to be shifted westward, as already identified for Kepler-7b by \cite{demory13}. Interestingly, Kepler-7b's high geometric albedo and day-side brightness temperature in the optical \citep{demory11, kipping11}, as well as the planet and host star radii \citep{latham10}, are all comparable to those of \kepb. However, Kepler-7b's mass is only $0.433 \pm 0.040$ \mjup, more than ten time less massive than \kepb\ mass, and its host star has a mass of about 1.35 \msun\ and effective temperature of about 6,000 K \citep{latham10}.

By analyzing the \ik\ phase curve we have identified a discrepancy between the beaming-based and ellipsoidal-based planet's mass estimates (see \tabr{planetmass} and \secr{planetmass}), which was already noticed before \citep{shporer11, mazeh12}. We chose to take a conservative approach and give a wide range for the planet's mass that includes both estimates.

If the early mid-occultation time is indeed due to the brightest region of the planetary atmosphere being shifted away from the substellar point then that may also affect the observed phase curve. Specifically, that will insert a phase shift to the reflection component in \eqr{beer} which includes both thermal emission and reflected stellar light from the planet's atmosphere. In that case the measured coefficients $a_{1c}$ and $a_{1s}$ would not correspond separately to the reflection and beaming amplitudes but a linear combination of them, meaning the model will include a degeneracy. This was already noted by \cite{faigler13} who analyzed the phase curve of Kepler-76b. Although the host star in that system is an F-star, so it is different than the host star in the \kepA\ system, they have shown that adding a phase shift to the reflection component results in a decreased beaming-based mass estimate for Kepler-76b and makes it consistent with the ellipsoidal-based mass estimate. Following a similar scheme \citep[see][ Eq.~1]{faigler13}, the required phase shift of the reflection component that will resolve the corresponding discrepancy for \kepb\ (see \tabr{planetmass}) is $1.49 \pm 0.48$ deg. However, such a phase shift will put the planet's brightest region eastward of the substellar point, so the resulting asymmetry in the occultation light curve ingress and egress will make the mid-occultation time later than expected, not earlier as we measure here. We conclude that if the measured early mid-occultation time is due to a planetary asymmetric surface brightness distribution we cannot detect direct evidence for it in the phase curve.

Our revised host star parameters show it is smaller than previous estimates \citep{szabo11}, leading to a smaller planetary radius of $R_p = 1.406 \pm 0.038$~\rjup\ where we assume the planet to star radii ratio reported by \cite{barnes11}. This revised planet radius is comparable with the radii of other hot Jupiters, although given the planet's relatively high mass it is positioned in a sparse region in the planetary radius-mass diagram for the currently known planets.

We have followed two approaches for interpreting our measurements and characterizing \kepb's atmosphere. In \secr{energybudget} we study the atmospheric energy budget, and in \secr{atm_model} we compare our wide-band measurements to various spectral atmospheric models. We must caution here that our conclusions about \kepb's atmosphere are based on sparse data. Although covering a wide wavelength range, from the IR to the optical, we have at hand the occultation depth measured in only four wide bands and the phase curve measured in only one wide band. Such sparse data could, in principle, lead to systematically biased results when fitted with over-simplified atmospheric models (see \citealt{burrows13} for a more detailed discussion). Despite the limited data, our analysis here, including the two different approaches (\secr{energybudget} and \secr{atm_model}), is an attempt to extract as much science as possible from the data while not over interpreting it. In the future, more detailed data, including panchromatic spectra and phase curves at various wavelengths, will allow a more comprehensive characterization of the planet's atmosphere.

The four occultation measurements enable us to identify the relation between the day-side brightness temperature and geometric albedo in each band (see \eqr{depth} and \figr{ag_td}). Assuming the geometric albedo does not change significantly between the four bands these relations allow us to derive the effective temperature of a black body that will show the same occultation depths, $\tdeff = 2,750 \pm 160$ K. This also results is a high geometric albedo, of $\ag = 0.33^{+0.04}_{-0.06}$, which assuming \ab = (3/2)\ag\ (Lambert's Law) leads to \ab = $0.50^{+0.06}_{-0.09}$. Such an albedo is at the high end of the range spanned by other hot Jupiters \citep[e.g.,][]{rowe08, cowan11, coughlin12, evans13, demory13, heng13}. 

The night-side brightness temperature in the optical is \tn = 2,537 $\pm$ 45 K, measured from the difference between the \ik\ occultation depth and the reflection component amplitude (see \eqr{ns}). \tn\ is smaller than \tdeff\ but not by much. Comparing \tn, \tdeff, and \ab\ results in inconsistencies, indicating the night side does not behave as a black body in the optical.

A possible explanation for the small difference between \tn\ and \tdeff\ is the planet's high mass, from at least five to nearly ten times larger than that of typical hot Jupiters, which gives a correspondingly large surface gravity with $\log_{10}(g\ [\rm g\ cm^{-2}])$ in the range of 3.79--4.01. Increased gravity leads to increased photospheric pressure which in turn increases the radiative time constant in the atmospheric layers probed by our measurements \citep{iro05, showman08}. As shown in idealized dynamical models by \cite{perez13}, atmospheres with greater radiative time constants exhibit smaller day-night thermal contrasts.

Comparing the wide-band occultation depths measured here to the spectral atmospheric models of \cite{fortney08} and \cite{burrows08} shows that our measurements are better described by models that include an atmospheric inversion and a weak day-night energy circulation. As can be seen in \figr{atm_model} the current atmospheric spectral models underestimate the occultation depth in the optical. Since the \ik\ occultation depth measurement is of much higher precision than occultation depths measured here in other bands, and than other occultations in the optical in other studies involving \ik\ data \citep[e.g.,][]{desert11a, desert11b, fortney11}, this discrepancy could be attributed to limits of the one-dimensional spectral models. If the \ik\ band measurement uncertainty was similar to that in the other bands then it would have been consistent with the models. Still, the fact that all models underestimate the \ik\ occultation depth suggests a higher geometric albedo in the optical than the typical 0.05 -- 0.10 predicted by the models. This supports the high geometric albedo of $\ag = 0.33^{+0.04}_{-0.06}$ derived in \secr{energybudget} for the equivalent black-body object showing the same occultation depths. If indeed \kepb\ day-side atmosphere has a high \ag\ then this could be used as a clue to identify the material dominating the day-side reflectivity, along with the day-side temperature and the possible existence of atmospheric inversion.

Short-period planets are expected to reach full orbital circularization and spin-orbit synchronization due to tidal interaction with the host star, an interaction that grows stronger with decreasing period \cite[e.g.,][]{mazeh08}. However, our measurement of the mid-occultation time suggests a possible small but finite orbital eccentricity (see \secr{timecomp}). If confirmed, it can lead to a non-synchronized planetary rotation \citep[e.g.,][]{correia11}, meaning the day and night sides are not permanent, which could explain the small brightness temperature difference detected between them in the optical. 

Our comparison between the atmospheres of \kepb\ and WASP-33b shows that despite the similarity between the host stars and other similarities between the two star-planet systems, the planetary atmospheres seem to be different. If confirmed, it is yet another example of the diversity of exoplanets and exoplanet atmospheres, emphasizing the need to discover more exoplanets that allow the study of their atmosphere, in particular those orbiting early-type stars like the one investigated here. Although spectra of such stars (currently) do not allow for high-precision RV measurements, preventing precise planet mass measurement \citep[but see][]{galland05, lagrange09}, the planetary nature of massive close-in planets can be confirmed with high-quality space-based photometry of bright stars. This was done using \ik\ data for \kepb\ and can potentially be done in the future with data from the NASA K2 mission \citep{howell14}, the NASA TESS mission\footnote{Scheduled for launch in 2017, see \url{http://tess.gsfc.nasa.gov}}, and the ESA PLATO mission \citep{rauer13}. In addition, detailed follow-up studies are also possible, like the measurement of stellar obliquity \citep{collier10} and investigation of the planet atmosphere. The host star's increased mass, radius, temperature, and younger age, compared to Sun-like stars, will allow testing planet formation and evolution theory.

%%%%%%%%%%%%%%%%%%%%%%%%%%%%%%%%%%%%%%%%%
\acknowledgments

% Shporer
A.S. thanks Ehud Nakar and Jason Eastman for enlightening discussions. 
This work was performed in part at the Jet Propulsion Laboratory, under contract with the California Institute of Technology (Caltech) funded by NASA through the Sagan Fellowship Program executed by the NASA Exoplanet Science Institute.
% O'Rourke
J.G.O. receives support from the National Science Foundation's Graduate Research Fellowship Program.
% Gyula Szabo
Gy.~M.~Sz. was supported by the Hungarian OTKA Grants 104607 and 83790, the HUMAN MB08C 81013 grant of the MAG Zrt and the J\'anos Bolyai Research Fellowship and a Lend\"ulet-2009 grant of the Hungarian Academy of Sciences.
% Ming Zhao
M.Z. is supported by the Center for Exoplanets and Habitable Worlds (CEHW) at the Pennsylvania State University. The Palomar/WIRC observation was in part supported by NASA through the American Astronomical Society's Small Research Grant program.
% ADS
This research has made use of NASA's Astrophysics Data System Service.
% Spitzer
This work is based on observations made by the {\it Spitzer Space Telescope}, which is operated by the Jet Propulsion Laboratory, California Institute of Technology, under a contract with NASA.
% P200
The Palomar 200 inch (5 m) Hale Telescope (P200) is operated by Caltech, JPL, and Cornell University.
% Kepler
Kepler was competitively selected as the tenth NASA Discovery mission. Funding for this mission is provided
by the NASA Science Mission Directorate. 
% Keck
Some of the data presented herein were obtained at the W.~M.~Keck Observatory, which is operated as a scientific
partnership among the California Institute of Technology, the University of California, and the National Aeronautics and Space Administration. The Keck Observatory was made possible by the generous financial support of the W.~M.~Keck Foundation.

%%%%%%%%%%%%%%%%%%%%%%%%%%%%%%%%%%%%%%%%%

%% Facilities
{\it Facilities: Warm \is, P200/WIRC, \ik, Keck/HIRES, P200/PHARO} 
%%%%%%%%%%%%%%%%%%%%%%%%%%%%%%%%%%%%%%%%%

%%%%%%%%%%%%%%%%%%%%%%%%%%%%%%%%%%%%%%%%%

\end{document}